\documentclass[superscriptaddress,amsmath,twocolumn,amssymb,prd,preprintnumbers,showpacs]{revtex4-2}
\usepackage{amssymb}
\usepackage{amsopn}
\usepackage{hyperref}
\usepackage{xcolor}
\usepackage[caption=false]{subfig}
\usepackage{soul}
\usepackage{amsmath}
\usepackage{graphicx}
\usepackage{float}
\usepackage{cleveref}
\usepackage[parfill]{parskip}
\usepackage{amsopn}
\usepackage{braket}
\usepackage{slashed}
\usepackage{bbold}

\begin{document}
\title{Perturbative S-matrix unitarity and higher-order Lorentz violation }
\author{Justo L\'opez-Sarri\'on}
\email[Electronic mail:]{ justo.lopezsarrion@ub.edu}
\affiliation{\it Departament de F\'\i sica Qu\`antica i Astrof\'\i sica and 
Institut de Ci\`encies del Cosmos (ICCUB), \\ Universitat de Barcelona, 
Mart\'i  Franqu\`es 1, 08028 Barcelona, Spain}
\author{Carlos M. Reyes}
\email[Electronic mail: ]{ creyes@ubiobio.cl}
\affiliation{ Centro de Ciencias Exactas, Universidad del B\'{i}o-B\'{i}o, 
 Casilla 447, Chill\'{a}n, Chile }
\author{ C\'esar Riquelme}
\email[Electronic mail:]{ ceriquelme@udec.cl}
\affiliation{ Centro de Ciencias Exactas, Universidad del B\'{i}o-B\'{i}o,
 Casilla 447, Chill\'{a}n, Chile }
\affiliation{ Departamento de F\' {\i}sica, Universidad de Concepci\'on, 
Casilla 160-C, Concepci\'on, Chile }
\begin{abstract}
We investigate the preservation of unitarity in a Lorentz and CPT-violating QED model
containing higher-order operators. In particular, we consider modifications in the fermion sector with 
dimension-five operators. The higher-order operators
 lead to an indefinite metric and a pseudo-unitarity relation for the $S$-matrix. 
However, we show that the pseudo-unitarity condition can be promoted to a genuine unitarity
 relation by i) restricting the energies to the effective region far below the Planck mass and 
 ii) considering stable particles to have positive metric. In the context of the optical theorem, 
 we focus on the one-loop Bhabha and Compton scattering processes. We show that no ghost 
 states get propagated through the cuts, thus satisfying the unitarity condition. Further, we show 
 that discontinuities of propagators are equivalent to replacing physical Dirac functionals in the cutting 
 equation. The physical Dirac functionals are defined 
 to select only mode solutions of stable particles. The provided extension of Cutkosky rule may be helpful
 for analyzing perturbative unitarity in higher-order diagrams.
\end{abstract}
\maketitle
\section{Introduction}
Over the last two decades, significant progress has been 
made in studying the possible breakdown of CPT and Lorentz invariance in extension to quantum field theories (QFT)
and gravity.
The combined efforts of theory, phenomenology, and ultra-high precision 
experiments have allowed to shape a robust effective framework
known 
as the Standard-Model Extension (SME) \cite{CK1, CK2}. The SME is an effective framework that accommodates
the most general parameterizations of CPT, local Lorentz, and diffeomorphism symmetry violations, extending both 
the standard model of particles and gravity.
The SME has established 
stringent limits on Lorentz violations and has identified the most promising sectors for detecting
 low-energy signatures of quantum gravity~\cite{Tables}. 
 
Extensions in QFT are typically achieved by introducing a privileged tensor that couples to both derivatives and fields.
The effective terms are kept small by a high degree of suppression of the Planck scale.
On the other hand, in gravity, the breaking of local Lorentz symmetry and diffeomorphism 
 has been more 
 searched with the mechanism of spontaneous symmetry breaking. In the case of spontaneous symmetry breaking,
 the background 
 fields acquire dynamics and introduce extra ingredients, such as massless 
 excitations or Nambu-Goldstone modes~\cite{Nambu:1960xd,Goldstone:1961eq}. The two 
 mechanisms have been called explicit and spontaneous symmetry breaking, respectively. 
In both cases, a background field with or without dynamics may arise by a non-trivial vacuum
in a more fundamental theory such as strings~\cite{strings,strings2}.

The effective field theories of the SME can be classified according to the mass dimensions of
 the operators introduced to describe Lorentz symmetry breaking. Specifically, they can be
divided into a 
 minimal sector with operators of mass dimensions up to four, and a nonminimal sector with 
higher-order operators.
Higher-order dimension operators have been a natural extension to include 
effects at higher energies in the effective framework. The exploration with 
 non-renormalizable operators are given in several sectors of the standard 
 model, photons~\cite{MKphotons}, fermions~\cite{MKfermions} and neutrinos~\cite{MKneutrinos}, 
 and also in linearized gravity~\cite{MKgravity}. 
 Several works study radiative corrections~\cite{Mariz:2018yvo,Cabral:2023xbf,Altschul:2022isc}, 
 vacuum Cherenkov radiation~\cite{Kaufhold:2005vj,Schreck:2017isa}, 
 explicit diffeomorphism breaking in gravity~\cite{ONeal-Ault:2020ebv,Reyes:2021cpx,Reyes:2022mvm}
 to mention some. 
 
 A potential drawback of higher-order operators 
 is that they may lead to
 the non-conservation of probability and to the loss of unitarity of the $S$-matrix~\cite{Pais:1950za}.
 However, it has been shown that there is no inherent contradiction in having 
 unitarity conserved, as certain prescriptions introduced by Lee and Wick can be followed~\cite{LW1,LW2}.
 The basic idea of the Lee-Wick prescription 
 is to restrict the asymptotic indefinite complex vector space, such to consider 
only positive metric particles to be stable.
 In this work, we discuss the main ingredients and the principal assumption under which an indefinite metric arises and how it
leads to the modification of the unitarity equation for the $S$-matrix, sometimes called pseudo-unitarity relation.
In recent years, many approaches 
have been developed to deal with the issue of unitarity conservation in Lorentz and CPT 
violating theories~\cite{Klinkhamer:2010zs,Reyes:2016pus,Ferreira:2020wde}.
The connection between the preservation of unitarity and Lorentz violation has been around for many years ago.
A few years after the works of Lee and Wick, Nakanishi pointed out that a modification in the contour of integration
 in Feynman diagrams may lead to the loss of covariance~\cite{Nakanishi:1971jj,Nakanishi:1972wx}. This may result 
 since momentum remains real, while in some regimes, the energy can become complex. Some further discussions on this context can be found in~\cite{Lee:1971ix}.

Recently the C-even part of the 
Myers and Pospelov model~\cite{MP} has been studied for testing perturbative unitarity at tree level~\cite{Lopez-Sarrion:2022czo}. 
Considering the Compton scattering process at tree level, it was shown that unitarity is preserved.
Here we extend the analysis to include the next natural step, which is to study perturbative unitarity at one-loop order.
To implement the generalization, we focus on two diagrams: the one-loop Bhabha and Compton scattering processes.
We take advantage and use
several expressions 
that have been derived, including
the dispersion relation, their mode, and eigenspinor solutions in~\cite{Lopez-Sarrion:2022czo}.

The organization of this work is as follows. In Sec.~\ref{sec:II}, we obtain the Myers-Pospelov timelike model 
starting from the generalized mass dimension fermion model of the SME. In Sec.~\ref{sec:III},
 we recall the dispersion relation, their mode
and spinor solutions that we have found previously in~\cite{Lopez-Sarrion:2022czo}. We discuss
 the interaction term and use it to compute 
the matrix elements of the $S$-matrix. Further, we provide a closed formula for the pseudo-unitarity 
relation in the presence of an indefinite metric.
 In Sec.~\ref{sec:IV}, we focus on the one-loop Compton and Bhabha scattering diagrams
 to study the preservation of unitarity. We use the perturbative tool of the optical theorem and check that 
 no ghost degrees of freedom are propagated through the cuts of amplitude diagrams.
 Sec.~\ref{sec:V} contains some further comments and a summary of our results.
\section{Modified fermion sector}\label{sec:II}
Our starting point is the
Lagrangian density of the 
fermion sector of the SME~\cite{MKfermions,MKneutrinos}
\begin{align}\label{Mod_fermions} 
\mathcal{L}_{\text{SME}}=\bar \psi (i \widehat \Gamma^{\mu}{\partial_{\mu}}-
\widehat{ M } ) \psi\,,
\end{align}
where all possible minimal and nonminimal contributions
 that break 
CPT and Lorentz symmetry can be expanded in terms of the $16$ Dirac matrices
\begin{subequations}
\begin{align}
\widehat \Gamma^{\mu}&=\gamma^{\mu}+\widehat c^{\alpha \mu}\gamma_{\alpha} 
+\widehat d^{\alpha \mu} \gamma_5\gamma_{\alpha}+\widehat e^{\mu}
 +i\widehat f^{\mu}\gamma_5 \notag \\ &\phantom{{}={}}+\frac{1}{2}
 \widehat g^{\kappa \lambda \mu} \sigma_{\kappa \lambda}     \,,
\end{align}
and
\begin{align}
\widehat M&=m+\widehat m+i\widehat m_5 \gamma_5+\widehat a^{\mu}
\gamma_{\mu}+\widehat b^{\mu}\gamma_5\gamma_{\mu}
\notag \\ &\phantom{{}={}} +\frac 12 \widehat H^{\mu \nu } \sigma_{\mu \nu}    \,,
\end{align}
\end{subequations}
where the effective derivative operators $\widehat c^{\alpha \mu}$, $\widehat d^{\alpha \mu}$, $\widehat m$,
$\widehat m_5, \widehat H^{\mu \nu }$ are CPT even, while
$\widehat e^{\mu}$ $\widehat f^{\mu}$ $\widehat g^{\kappa \lambda \mu}$
$\widehat a^{\mu}$, $\widehat b^{\mu}$ are CPT odd.

We are interested in making
the connection with 
the Myers and Pospelov (MP) model~\cite{MP}, which contain
dimension-five operators. Hence, we turn off several effective terms and retain
\begin{subequations}
\begin{align} \label{Gamma}
\widehat \Gamma^{\mu}&=\gamma^{\mu} \,, \\ \label{M}
 \widehat{M } &= m+\widehat a^{(5)\mu}\gamma_{\mu}+\widehat b^{(5)\mu} \gamma_5 \gamma_{\mu} \,,
\end{align}
with
\begin{align}\label{a}
\widehat a^{(5)\mu}&:= - \frac{\eta_1}{m_{\textrm{Pl} }}(n\cdot
\partial)^2n^{\mu} \,, \\ \label{b}
 \widehat b^{(5)\mu}&:=  \frac{\eta_2}{m_{\textrm{Pl} }} (n\cdot 
 \partial)^2n^{\mu} \,.
\end{align}
\end{subequations}
Considering the operators~\eqref{Gamma} and \eqref{M}  and replacing
in equation~\eqref{Mod_fermions}, 
we arrive at the fermion MP Lagrangian density
\begin{align} \label{lagMP}
\mathcal{L}_{\text{MP}}&=\bar \psi (i \slashed{\partial}-m ) \psi  
+\frac{ \bar
 \psi   } {m_{\textrm{Pl}} }\left(\eta_1 \slashed{n}+\eta_2 \slashed{n} \gamma_5 \right)\nonumber
   \\    &\phantom{{}={}}\times
    (n\cdot \partial  )^2  \psi  \,.
\end{align}
where $n^{\mu}$ is a
preferred four-vector responsible to break Lorentz symmetry, $m_{\textrm{Pl}}$
is the Planck mass and $\eta_1$, $\eta_2$ are coupling constants. Also one can show that  
$\eta_1$ is charge conjugation odd and $\eta_2$ charge conjugation even.

The free equation of motion is 
\begin{equation} \label{free_motion}
\left( i \slashed{\partial}-m   +\frac{  1  } {m_{\textrm{Pl}} } 
(\eta_1 \slashed{n}+\eta_2 \slashed{n} \gamma_5)
    (\partial \cdot n  )^2 \right) \psi(x) =0 \,.
\end{equation}
Using the plane wave ansatz $\psi(x)=\int d^3k \,\psi(k) \,e^{-ik\cdot x}$
 the dispersion relation becomes
\begin{align}\label{Disp_Rel}
  & \left( p^2-m^2-2g_1 \left(n\cdot p\right)^3 +  
   n^2 (g_1^2-g_2^2) (n\cdot p)^4 \right)^2\notag \\
   &\phantom{{}={}}-4g_2^2 \left(n\cdot p\right)^4 D(n,p)=0 \,,
\end{align}
where we define $g_1:=\frac{\eta_1}{m_{\textrm{Pl}}}$, 
$g_2:=\frac{\eta_2}{m_{\textrm{Pl}}}$
and $D(n,p):= (n\cdot p)^2-p^2n^2 $.

In this work we utilize the chiral representation
for Dirac matrices, i.e,
\begin{equation}
\gamma^\mu =\left( \begin{array}{c c}
0\quad & \sigma^\mu\\
\bar\sigma^\mu  \quad &0
\end{array}\right)\,, \qquad 
\gamma_5=\left(\begin{array}{c c}
-\mathbb{1}_{2} \quad &0\\
0 \quad &\mathbb{1}_{2} \end{array}\right)\,,
\end{equation}
with
$\sigma^\mu=(\mathbb{1}_{2},\vec\sigma)$,
$\bar\sigma^\mu=(\mathbb{1}_{2},-\vec\sigma)$ and 
$\mathbb{1}_{2}$ the $2\times2$ identity matrix.
For the metric signature in Minkowski spacetime
we employ the mostly minus sign convention $(+,-,-,-)$.
\section{The timelike model}\label{sec:III}
In this section, we provide the basic properties
 of the fermion timelike MP model. We take advantage of the dispersion relations, their modes solutions, and eigenspinors
presented in~\cite{Lopez-Sarrion:2022czo}.
We additionally discuss the interaction term in our modified QED model 
and present a detailed derivation of
the unitary equation satisfied by 
the $S$-matrix in an indefinite metric theory.
\subsection{Dispersion relation and spinors}\label{subsec:III-1}
We set the background four-vector in the pure timelike
direction $n^{\mu}=(1,0,0,0)$ and turn off the charge conjugation odd sector with $\eta_1=0$.
With these choices, the Lagrangian~\eqref{lagMP} takes the form
\begin{align} \label{lagMPEta=0}
  \mathcal{L}_{\substack{\text{MP} \\ \text{C-even}    }} &=\bar \psi (i \slashed{\partial}-m ) \psi  +g_2\bar \psi \gamma_0 \gamma_5  \ddot{\psi} \,.
\end{align}	
The dispersion relation can be
 found from Eq.~\eqref{Disp_Rel} to be
\begin{align}\label{SimpDispersion}
	\left(p_0^2-  |\vec{p}|^2-m^2-g_2^2 p_0^4 \right)^2-4g_2^2 p_0^4|\vec{p}|^2=0\,.
\end{align}

Let us introduce the quantities
\begin{subequations}
\begin{align}\label{lambdas}
 \Lambda_{+}^2 (p)&=: p_0^2-|\vec{p}|^2 -m^2 -   
g_2^2 p_0^4-2 g_2p_0^2 |\vec{p}|\,,
\end{align}
 \begin{align}\label{lambdas2}
 \Lambda_{-}^2(p) &=: p_0^2-|\vec{p}|^2 -m^2 - g_2^2 p_0^4+2 g_2p_0^2|\vec{p}|\,.
\end{align}
\label{lambdas,global}
\end{subequations}
In terms of these quantities
the dispersion relation can be written as
\begin{align}\label{SimpDispersion}
\Lambda_+^2(p)	 \Lambda_-^2(p)& = 
\left(p_0^2-  |\vec{p}|^2-m^2-g_2^2 p_0^4 \right)^2
\notag \\ &\phantom{{}={}} -4g_2^2 p_0^4|\vec{p}|^2\,.
\end{align}

The solutions of the equation
$\Lambda_+^2(p)=0$  are
\begin{subequations}
\begin{align} \label{posit_freq1}
	\omega_1&=\sqrt{\frac{1-2g_2 |\vec p |- \sqrt{(1-2g_2 |\vec p |  )^2 -
	4g_2^2E_p^2}}{2g_2^2} }\,,   \notag
  \\
	W_1&=\sqrt{\frac{1-2g_2 |\vec p | + \sqrt{(1-2g_2 |\vec p |  )^2 -
	4g_2^2E_p^2}}{2g_2^2} }\,,  
\end{align}
and the solutions of the equation 
$\Lambda_-^2(p)=0$ 
\begin{align} \label{posit_freq2}
	 \omega_2&=\sqrt{\frac{1+2g_2 |\vec p |- \sqrt{(1+2g_2 |\vec p |  )^2
	 -4g_2^2E_p^2}}{2g_2^2} }   \,,	\notag  \\
	  W_2&=\sqrt{\frac{1+2g_2 |\vec p | + \sqrt{(1+2g_2 |\vec p |  )^2 
	-4g_2^2E_p^2}}{2g_2^2} }   \,.
\end{align}
\end{subequations}
We also have the negative mode solutions $-\omega_1, -W_1, -\omega_2, -W_2$,
where $E_p=\sqrt{ |\vec p |^2+ m^2}$. In~\cite{Lopez-Sarrion:2022czo},
we have shown that modes $\pm W_{1,2}$
correspond to heavy ghost modes while $\pm \omega_{1,2}$ can be associated with perturbative 
modes of standard particles.

In section~\ref{sec:IV}, we study the unitarity of the model and will need to examine the modes in the complex $p_0$-plane. The poles $\omega_1$ and $W_1$ exhibit 
a peculiar momentum-dependent behavior. As the magnitude of spatial momenta $|\vec{p}|$ increases, $\omega_1$ moves in the positive direction of the real axis, while $W_1$ moves 
in the opposite direction. The two modes coincide at
\begin{align}\label{pmax}
\vert\vec p\vert_{\text{max}}=\frac{1-4g_2^2m^2}{4g_2}\,,
\end{align}
 where they both take the 
value $\frac{1}{2g_2}\sqrt{1+4g_2^2m^2}$, and for higher momenta, they start to have an imaginary component. For complex $\omega_1$ and 
$W_1$, the first solution moves downward in the imaginary axis, while the latter heavy mode $W_1$ moves upward, see Fig.~\ref{Fig2}. The solutions $\omega_2$ 
and $W_2$ remain real for all momentum values. The negative modes behave similarly, with the only difference being that when they become 
complex, the solution $-\omega_1$ moves upward, while $-W_1$ moves downward.

We can write the dispersion relation~\eqref{SimpDispersion} as
\begin{align}
		 \Lambda_+^2(p)	 \Lambda_-^2(p) &=  g_2^4  (p_0^2-\omega_1^2)    (p_0^2-W_1^2)
		  (p_0^2-\omega_2^2) \notag \\ &\phantom{{}={}} \times (p_0^2-W_2^2)=0\,.
\end{align}

The positive-energy eigenspinors for $\Lambda_+^2(p)=0$ are
\begin{subequations}
\begin{align}
	u^{(1)}( p)&=\left(\begin{array}{c}   \sqrt{p_0-g_2p_0^2-|\vec{p}| } \xi^{(+)}(\vec p)\\ 
	 \sqrt{p_0+g_2p_0^2+|\vec{p}| }  \xi^{(+)}(\vec p) \end{array}\right)_{p_0=\omega_1}  \,,
 \\
	U^{(1)}( p)&=\left(\begin{array}{c}  \sqrt{p_0-g_2p_0^2-|\vec{p}| }\xi^{(+)}(\vec p)\\ 
	\sqrt{p_0+g_2p_0^2+|\vec{p}| }\xi^{(+)}(\vec p) \end{array}\right)_{p_0=W_1}  \,,
\end{align}
\label{spinors1}
\end{subequations}
and for $\Lambda_-^2(p)=0$ 
\begin{subequations}
\begin{align}
	u^{(2)}( p)&=\left(\begin{array}{c} \sqrt{p_0-g_2p_0^2+|\vec{p} |}  \xi^{(-)}(-\vec p)\\ 
	\sqrt{p_0+g_2p_0^2-|\vec{p} |}  \xi^{(-)}(-\vec p) \end{array}\right)_{p_0=\omega_2}  \,,
 \\
	U^{(2)}( p)&=\left(\begin{array}{c}  \sqrt{p_0-g_2p_0^2+|\vec{p} |}  \xi^{(-)}(-\vec p)\\ 
	\sqrt{p_0+g_2p_0^2-|\vec{p} |} \xi^{(-)}(-\vec p) \end{array}\right)_{p_0=W_2}  \,.
\end{align}
\label{spinors2}
\end{subequations}
The negative-energy eigenspinors for $\Lambda_+^2(p)=0$ are
\begin{subequations}
\begin{align}
	v^{(1)}( p)&=\left(\begin{array}{c}   \sqrt{p_0+g_2p_0^2+|\vec{p}| } \xi^{(-)}(-\vec p)\\ 
	 - \sqrt{p_0-g_2p_0^2-|\vec{p}| }  \xi^{(-)}(-\vec p) \end{array}\right)_{p_0= \omega_1}  \,,
 \\
	V^{(1)}( p)&=\left(\begin{array}{c}  \sqrt{p_0+g_2p_0^2+|\vec{p}| }\xi^{(-)}(-\vec p)\\ \label{v_1}
	- \sqrt{p_0-g_2p_0^2-|\vec{p}| }\xi^{(-)}(-\vec p) \end{array}\right)_{p_0= W_1}  \,,
\end{align}
\end{subequations}
and for $\Lambda_-^2(p)=0$ 
\begin{subequations}
\begin{align}
	v^{(2)}(p)&=\left(\begin{array}{c} \sqrt{p_0+g_2p_0^2-|\vec{p} |}  \xi^{(+)}(\vec p)\\ 
	- \sqrt{p_0-g_2p_0^2+|\vec{p} |}  \xi^{(+)}(\vec p) \end{array}\right)_{p_0= \omega_2}  \,,
 \\
	V^{(2)}( p)&=\left(\begin{array}{c}  \sqrt{p_0+g_2p_0^2-|\vec{p} |}  \xi^{(+)}(\vec p)\\ \label{v_2}
-	\sqrt{p_0-g_2p_0^2+|\vec{p} |} \xi^{(+)}(\vec p) \end{array}\right)_{p_0= W_2}  \,.
\end{align}
\end{subequations}

Above we have introduced the bi-spinors $\xi^{(\pm)}(\vec p)$,
given by
\begin{align}
\xi^{(+)}(\vec p)&= \frac{1}{\sqrt{  2|\vec p| \left(|\vec p|+p^3 \right)}}   \left(\begin{array}{c}
|\vec p| +p^3 	\\
p^1+ip^2	
\end{array}\right) \,,
\\
\xi^{(-)}(\vec p)&= \frac{1}{\sqrt{  2|\vec p| (|\vec p|-p^3)}}   \left(\begin{array}{c}
p^1-ip^2		\\  |\vec p| -p^3
\end{array}\right) \,,
\end{align}
which satisfies the properties
\begin{align}\label{propc}
	 (\vec{p}\cdot\vec{\sigma}  )  \xi^{(\pm)}  (\vec p) &= |\vec p| \xi^{(\pm)}(\vec p) \,,
\notag \\
	 (\vec{p}\cdot\vec{\sigma}  )  \xi^{(\pm)}  (-\vec p) &=- |\vec p| \xi^{(\pm)}(-\vec p) \,,
\end{align}
and the orthogonality relations 
\begin{align}\label{Ort_prop_bi}
\xi^{(+) \dag}(\vec p)  \xi^{(+)}(\vec p)&=\xi^{(-) \dag}(\vec p)  \xi^{(-)}(\vec p)=1\,,
\notag \\
\xi^{(+) \dag }(\vec p)  \xi^{(-)}(-\vec p)&=\xi^{(-) \dag }(-\vec p)  \xi^{(+)}(\vec p)=0\,,
\end{align}
together with
\begin{align}\label{ep1}
 \xi^{(\pm)  }(\vec p)  \xi^{(\pm) \dag }(\vec p)&=\frac{1}{2}\left(1+\frac{\vec \sigma \cdot \vec p} {|\vec p|} \right)  \,,
 \\ \label{ep2}
   \xi^{(\pm)  }(-\vec p)  \xi^{(\pm) \dag }(-\vec p)&=\frac{1}{2}\left(1-\frac{\vec \sigma \cdot \vec p} {|\vec p|} \right)\,.
\end{align}

The modified propagator is found to be
\begin{align}\label{S_Fprop}
S_F(p)&=\frac{i  F(p)}{ D_p } \,,
\end{align}
where 
\begin{align}
F(p)&=\bar{M}(p)N(p)\bar{N}(p)   \,,    \\    D_p &= \Lambda_+^2
(p+i\epsilon)  \Lambda_-^2 (p+i\epsilon)    \,,
\end{align}
with
 \begin{subequations}
 \begin{align}
	 M&=\slashed{p}-m-g_2p_0^2  \gamma_0  \gamma_5 \,,  \label{Op_M}\\
	\bar {  M}&=\slashed{p}+m-g_2 p_0^2 \gamma_0\gamma_5  \label{Op_Mbar} \,,\\
	 { N}&=\slashed{p}+m+g_2 p_0^2 \gamma_0\gamma_5  \,, \label{Op_N}\\   \label{Op_Nbar}
  \bar {N}&=\slashed{p}-m+g_2p_0^2 \gamma_0\gamma_5 \,.
\end{align}
\end{subequations}
By taking into consideration the $i\epsilon$ prescription
in~\eqref{lambdas} and~\eqref{lambdas2} 
the pole structure is described by
 \begin{subequations}
\begin{align}
	 \Lambda_+^2(p+i\epsilon)&=-g_2^2(p_0+\omega_1-i\varepsilon) 
  (p_0-\omega_1+i\varepsilon) \notag  \\ &\phantom{{}={}}\times 
(p_0+W_1-i\varepsilon) (p_0-W_1+i\varepsilon)  \,,
 \\
  \Lambda_-^2 (p+i\epsilon)&=-g_2^2 (p_0+\omega_2-i\varepsilon) 
  (p_0-\omega_2+i\varepsilon)   
    \notag  \\   &\phantom{{}={}}\times    (p_0+W_2-i\varepsilon)
    (p_0-W_2+i\varepsilon)   \,.
\end{align}
\end{subequations}
Note that negative and positive solutions are placed above and below the real axis, respectively, in the 
complex $p_0$-plane. We also have the identities
\begin{subequations}
\begin{align} \label{disprelaid1}
2\left(p^2-m^2-g_2^2 p_0^4\right)_{p_0=\omega_1} &=-g_2^2(\omega_1^2-\omega_2^2) (\omega_1^2-W_2^2) \,,
   \\  \label{disprelaid2}
2\left(p^2-m^2-g_2^2 p_0^4\right)_{p_0=\omega_2} &=-g_2^2(\omega_2^2-\omega_1^2)   (\omega_2^2-W_1^2)  \,.
\end{align}
\end{subequations}
In section~\ref{sec:IV} we use the expressions 
\begin{subequations}
\begin{align} 
u^{(1)}( p)\bar{u}^{(1)}( p)&=\frac{F(\omega_1,\vec p)}{g_2^2
(\omega_1^2-\omega_2^2)(W_2^2-\omega_1^2)}  \,. \label{u1u1} \\
u^{(2)}( p)\bar{u}^{(2)}( p)&=-\frac{F(\omega_2,\vec p)}{g_2^2
(\omega_1^2-\omega_2^2)(W_1^2-\omega_2^2)} \,, \label{u2u2}\\
v^{(1)}(p)\bar{v}^{(1)}( p)&=-\frac{F(-\omega_1,-\vec p)}{g_2^2
(\omega_1^2-\omega_2^2)(W_2^2-\omega_1^2)}  \,, \label{v1v1}\\
v^{(2)}(  p)\bar{v}^{(2)}( p)&=\frac{F(-\omega_2,-\vec p)}{g_2^2
(\omega_1^2-\omega_2^2)(W_1^2-\omega_2^2)}  \,. \label{v2v2}
\end{align}
\end{subequations}
The demonstration of the above relations are 
not difficult, however, we proceed to prove the first one with the other ones
following similarly. 
From~\eqref{Op_N} and~\eqref{Op_Nbar} we find
\begin{align}
N\bar N=p^2-m^2-g_2^2p_0^4+2g_2p_0^2p_i\gamma^i\gamma_0\gamma_5\,,
\end{align}
and evaluated in the mode $\omega_1$
\begin{align}
\left(N\bar N\right)_{p_0=\omega_1}=  2g_2 \omega_1^2 |\vec p |    \left( 1+\frac{ p_i \gamma^i\gamma_0\gamma_5}{ |\vec p | } \right ) \,.
\end{align}
From~\eqref{ep1} we have
\begin{align}\label{NbarN}
\left(N\bar N\right)_{p_0=\omega_1}=  4g_2 \omega_1^2 |\vec p |  \;   \xi^{(+)  }(\vec p)  \xi^{(+) \dag }(\vec p) \,.
\end{align}
On the other hand, working directly with the eigenspinors, we can prove that
\begin{align}
u^{(1)}(p)\bar{u}^{(1)}(p)=(M)_{p_0=\omega_1} \xi^{(+)  }(\vec p)  \xi^{(+) \dag }(\vec p)\,,
\end{align}
and hence replacing~\eqref{NbarN} above and
using
 $(p^2-m^2-g_2^2p_0^4)_{p_0=\omega_1}=2g_2 \omega_1^2 |\vec p |  $ and~\eqref{disprelaid1}, we have the relation 
\begin{align}
u^{(1)}(p)\bar{u}^{(1)}(p)=\frac{(M N\bar N)_{p_0=\omega_1} }{g_2^2(\omega_1^2-\omega_2^2   )   (W_2^2-\omega_1^2   )  }\,.
\end{align}

A detailed derivation of the previous expressions, including the eigenspinors, the propagator and the consistency of
the pole prescription
can be found in~\cite{Lopez-Sarrion:2022czo}.
\subsection{The QED model}\label{subsec:III-2}
Consider the higher-order Lorentz-violating QED that arises
by 
performing the minimal 
coupling substitution in the modified MP fermion
Lagrangian~\eqref{lagMP} and coupling to the Maxwell Lagrangian
\begin{align} \label{TotalLagrangian}
\mathcal{L}_{\text{QED}}&=\bar \psi (i \slashed{D}-m ) \psi  +\frac{1}{m_{\textrm{Pl}}}\bar
 \psi (\eta_1 \slashed{n}    +\eta_2 \slashed{n} \gamma_5)
   (D \cdot n)^2   \psi  \notag \\ &\phantom{{}={}} -\frac{1}{4} F_{\mu\nu}F^{\mu\nu} \,, 
\end{align}
where $D_\mu=\partial_\mu+ieA_\mu$ and 
 $F_{\mu \nu}=\partial_{\mu} A_{\nu}-\partial_{\nu} A_{\mu}$ is the usual field strength tensor.
The local gauge invariance of the Lagrangian~\eqref{TotalLagrangian}
can be checked
by using the gauge transformations of the fields
\begin{subequations}
 \begin{align}
 A'_{\mu}(x)&=A_{\mu}(x)+ \partial_{\mu}\lambda(x)\,,  \\  
 \psi' (x)&=e^{-ie\lambda(x)} \psi (x)\,,
 \end{align}
and the induced 
 transformations for the derivative terms
 \begin{align}
D_{\mu} \psi' &= e^{-i e \lambda}D_{\mu}  \psi \,,
\\
D_{\alpha}  D_{\mu} \psi' & = e^{-i e \lambda} D_{\alpha}D_{\mu}  \psi  \,.
\end{align}
\end{subequations}
We can always choose to work in
the axial gauge
where the gauge field is imposed to fulfill the relation $A\cdot n=0$. This choice is advantageous, since 
we arrive at 
 the usual QED interaction, and so is the one we use throughout this work. Considering the restricted C-even part
  and timelike fermion sector
 produces the modified QED Lagrangian 
\begin{align} \label{Lagtimelike}
\mathcal{L'}_{\text{QED}}&=\bar \psi (i \slashed{\partial}+e \slashed{A}  -m ) 
\psi  +g_2\bar \psi \gamma_0 \gamma_5  \ddot{\psi} \notag \\  &\phantom{{}={}}
  -\frac{1}{4} F_{\mu\nu}F^{\mu\nu} \,.
\end{align}
\subsection{Indefinite metric theories }\label{subsec:III-3}
Here we briefly discuss the general aspects of the unitarity 
of theories that have regular and indefinite metric $\eta$.

Consider a complex vector space $\mathcal F$ with vector basis  
 $\{ \vert\alpha \rangle\} \in \mathcal B$ and metric
 \begin{align}
\eta_{\alpha\beta} :=\langle\alpha\vert\beta\rangle       \,.
 \end{align}
We assume the metric not to be positive definite, so in principle each element $\eta_{\alpha\beta}$
may have a positive or negative value.

A representation of the identity operator is
\begin{equation}\label{unity}
\mathbb{I}=\sum_{\alpha,\beta\in\mathcal{B}} \vert\alpha
\rangle  \, \eta^{-1}_{\alpha\beta}  \,\langle\beta\vert\,. 
\end{equation}
The matrix elements of an arbitrary linear operator $\mathcal{U}$, are defined as, 
$\widetilde{U}_{\alpha\beta}  := \langle\alpha\vert \mathcal{U}\vert\beta\rangle$  
and the unitarity condition for $\mathcal{U}$  is the requirement that 
the  linear transformation leaves the inner product invariant, i.e.,
 \begin{align}
\langle\alpha\vert\beta\rangle=\langle \mathcal{U}\alpha\vert 
\mathcal{U}\beta\rangle=\langle\alpha\vert\mathcal{U}^\dag \mathcal{U}\vert\beta\rangle\,,
 \end{align}
or in terms of the matrix elements written as~\cite{LW1,LW2}
 \begin{align}
\eta_{\alpha\beta}&=\sum_{\alpha^\prime,\beta^\prime}\langle\alpha\vert 
\mathcal{U}^\dag\vert\alpha^\prime\rangle
\eta^{-1}_{\alpha^\prime\beta^\prime}\langle\beta^\prime\vert 
\mathcal{U}\vert\beta\rangle \notag \\ &  =\sum_{\alpha^\prime,\beta^\prime} 
\widetilde{U}^\dag_{\alpha\alpha^\prime}
\eta^{-1}_{\alpha^\prime\beta^\prime}\widetilde{U}_{\beta^\prime\beta}\,,
 \end{align}
where we have used (\ref{unity}).  
In matrix notation, we have the expression
\begin{equation}
\widetilde{U}^\dag\eta^{-1}\widetilde{U}=\eta\label{pseudoUnitarity}\,,
\end{equation}
for the pseudo-unitarity condition 
in the presence of
an indefinite metric $\eta$.
For a theory with indefinite metric 
we should expect to have an evolution
operator $U$ to have this property. However, the inner product cannot be interpreted 
as a probability amplitude, as it can only have a meaningful interpretation in the positive metric sector.
Then, if $\mathcal{U}$ stands for the time evolution, we can write,
 \begin{align}
\mathcal{U}=\mathcal{I}+i\mathcal{T}\,,
 \end{align}
or in terms of matrix  elements, by projecting the previous 
equation on $\langle\alpha\vert$ and $\vert \beta\rangle$
 \begin{align}
\widetilde{U}_{\alpha\beta}=\eta_{\alpha\beta}
+i\widetilde{T}_{\alpha\beta}\,,
 \end{align}
where the $\widetilde{T}$ is the transition matrix 
(or $S$ matrix) for a dynamically evolving  state.  
So, the condition of the invariance of the inner product becomes, 
in terms of matrix elements
 \begin{align}
-i(\widetilde{T}  - \widetilde{T}^\dag)
=\widetilde{T}^\dag \eta^{-1} \widetilde{T}\,.
 \end{align}
The diagonal part of this matrix equation can be written as
 \begin{align}
2 \text {Im}(\widetilde{T} )_{\alpha\alpha}=
(\widetilde{T}^\dag\eta^{-1} \widetilde{T})_{\alpha\alpha}\,,
 \end{align}
which is the pseudo-unitarity version of the optical theorem.  

Now, let us write all in terms of a basis build-up of physical particles 
and ghost states, where any state of the basis is of the form
 \begin{align}
\vert \alpha\rangle &= \vert p_1,r_1\rangle
\otimes\vert p_2,r_2\rangle \otimes\cdots\otimes
\vert p_N,r_N\rangle \notag \\  &\phantom{{}={}} \otimes 
\vert \widetilde{q}_1,s_1\rangle\otimes\vert \widetilde{q}_2,s_2\rangle
\otimes \cdots\otimes\vert \widetilde{q}_M,s_M\rangle\,,
 \end{align}
where $p_i,r_i$ are momentum and other  quantum numbers of physical 
particles and $q_j,s_j$ are momentum and other quantum numbers of 
ghost.
Impose one-particle and ghost states being orthogonal,
 \begin{align}
\langle p_i,r_i\vert p_j,r_j\rangle =N_i \delta^3(p_i-p_j)\delta_{r_ir_j}\,,
 \end{align}
 \begin{align}
\langle \widetilde{q}_i,s_i\vert \widetilde{q}_j,s_j\rangle =-\mathcal{N}_i 
\delta^3(  \widetilde q_i-  \widetilde q_j)\delta_{s_is_j}\,,
 \end{align}
and
 \begin{align}
\langle p_i,r_i\vert {\widetilde q_j,s_j}\rangle =0\,,
 \end{align}
with $N_i$ and $\mathcal{N}_i$ positive normalization constants. From this 
basis, it is easy to build up the matrix elements of $\eta$. The  choice of 
$N_i=\mathcal{N}_i=1$ simplifies the expressions because in this case we 
have that $\eta_{\alpha\alpha}=\pm1$ and then, $\eta^{-1}=\eta$. In the 
rest of this section, we use this choice for simplicity.

Following the Lee-Wick prescription, only states with particles will be considered 
as asymptotic states. So, it is convenient to split the space of states, 
$\mathcal{F}$, in two orthogonal subspaces,
$$\mathcal{F}=\mathcal{V}^+\oplus\mathcal{V}^-$$ 
The first one, The   sector spanned by the physical particle states, which we call the
physical sector, ${\cal V}^+$, generated by the basis $\mathcal{B}^+$
 \begin{align}
\mathcal{B}^+=\left\{\vert p_1,r_1\rangle\otimes\vert p_2,r_2
\rangle \otimes\cdots\otimes\vert p_N,r_N\rangle\right\}_{p_i,r_i,M}\,,
 \end{align}
And the unphysical space, $\mathcal{V}^-$, its orthogonal 
complement, which is spanned by $\mathcal{B}^-$, given by ghost particle states.

The interpretation of probability amplitudes of the inner product is 
meaningful only for the physical sector $\mathcal{V}^+$. So, pseudo-unitarity 
of time evolution 
is compatible with probability conservation if 
restricted to the physical sector one has the standard unitarity relation
 \begin{align}
2\text{Im} \left(\langle\mathrm{phys}\to \mathrm{phys}\right)=\lvert 
\langle\mathrm{phys}\to \mathrm{any \,phys}\rangle\rvert^2\,.
 \end{align}
Our pseudo-unitarity condition, however, restricted to physical 
state $\alpha\in \mathcal{V}^+$ is
 \begin{align}\label{CE}
2\text{Im} (T_{\alpha \to\alpha})=\sum_{\beta\in\mathcal{B}^+} \lvert 
T_{\alpha\to\beta}\rvert^2+ \sum_{\gamma\in\mathcal{B}^-} 
\eta_{\gamma\gamma}\lvert  T_{\alpha\to\gamma}\rvert^2\,,
 \end{align}
which is just the standard optical theorem if the last term vanishes. 
So the conclusion is that probability is conserved if and only if the
transitions between physical states and nonphysical states always vanish. 
In this case, the probability is well defined and unitarity 
can be realized in the theory.
\section{One-loop diagrams }\label{sec:IV}
As demonstrated in the previous section, the physical $S$-matrix evolves
 unitarily if the last term in Eq.\eqref{CE} is zero. In the following section, we investigate
  whether this condition can be satisfied in our modified QED~\eqref{Lagtimelike}. We focus 
  on the two relevant $2\to 2$ scattering processes being
   the one-loop Bhabha and Compton reactions.
\subsection{One-loop Bhabha scattering}\label{subsec:IV-1}
\begin{figure}
    \centering
    \includegraphics[scale=0.48]{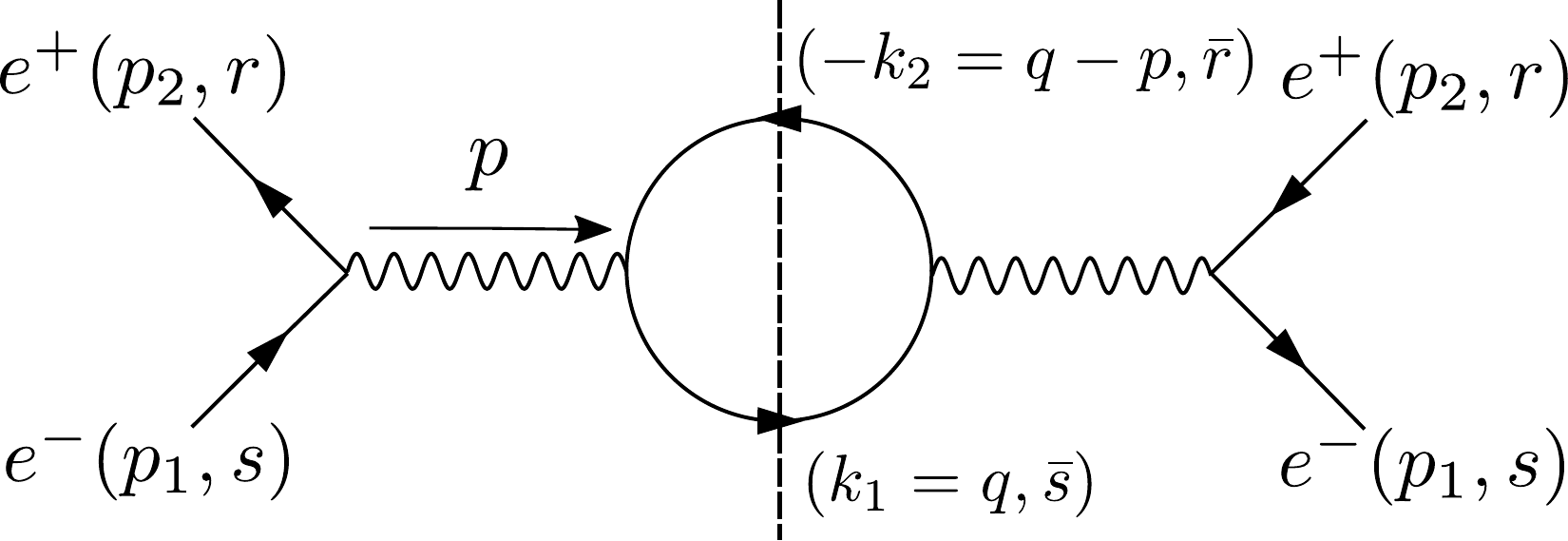}
    \caption{ \label{Fig1} The one-loop Bhabha scattering 
    process $e^+e^-\to e^+e^-$ and the cut diagram of the RHS of equation~\eqref{CE} 
    indicated with the vertical segmented line.}
\end{figure}

We start with the electron and positron forward 
scattering reaction $e^-(p_1,s)+e^+   (p_2,r)  \to e^-(p_1,s)+e^+   (p_2,r)$, see Fig.~\ref{Fig1}.

The amplitude is given by
\begin{align}
i\mathcal{M}^{(1)}_F&=(-1)  \left( \bar{v}^r(p_2)(-ie\gamma^\mu)u^s(p_1) \right)
\left(\frac{-i\eta_{\mu\nu}}{p^2+i\epsilon}\right) \nonumber 
\\ &\phantom{{}={}} \times \int \frac{d^4 q}{(2\pi)^4}\text{Tr}
\left[(-ie\gamma^\rho)S_F(q-p)(-ie\gamma^\nu) S_F(q)\right]\notag \\ 
& \phantom{{}={}}\times\left(\frac{-i\eta_{\rho\sigma}}{p^2+i\epsilon} 
\right)\left(\bar{u}^s(p_1)(-ie\gamma^\sigma)v^r(p_2) \right)\,.
\end{align} 
In terms of the propagator in~\eqref{S_Fprop}, and the currents
\begin{subequations}
\begin{align}
   & J_1^\mu(p_1,p_2):=\bar{v}^r(p_2)\gamma^\mu u^s(p_1)\,,
\\
    &J_2^{\mu}(p_1,p_2):=\bar{u}^s(p_1)\gamma^\mu v^r(p_2) =[J_1^{\mu}(p_1,p_2)]^*\,,
\end{align}
\end{subequations}
we write
\begin{align}\label{Amp2}
 \mathcal{M}_F ^{(1)}  &=\frac{i e^4}{(p^2+i\epsilon)^2} J_1^\mu(p_1,p_2) J_2^\nu(p_1,p_2)  \\  \notag &
 \phantom{{}={}}\times \int \frac{d^3 \vec q}{(2\pi)^4} \oint_{\mathcal C^{(1)}} dq_0 \,
 \text{Tr}\left[\gamma_\nu \frac{F(q-p)}{D_{q-p}}\gamma_\mu \frac{F(q)}{D_q}\right]\,.
\end{align}

The poles in complex $q_0$-plane are dominated by the quantities
\begin{subequations}
\begin{align}
 \frac{1}{D_q}&= \frac{1}{g_2^4} \prod_{s=1,2}\Bigg[ \frac{1}{(q_0+\omega_s-i\epsilon)
 (q_0-\omega_s+i\epsilon) }\notag \\ 
  &\phantom{{}={}}\times \frac{1}{(q_0+W_s-i\epsilon) (q_0-W_s+i\epsilon) }  \Bigg]\,,
\end{align}
and
\begin{align}
 \frac{1}{D_{q-p}}&= \frac{1}{g_2^4} \prod_{s=1,2}
 \Bigg[ \frac{1}{(q_0-p_0+\overline {\omega}_s-i\epsilon)
 (q_0-p_0-\overline { \omega}_s+i\epsilon) }\notag \\ 
  &\phantom{{}={}}\times \frac{1}{(q_0-p_0+\overline { 
  W}_s-i\epsilon) (q_0-p_0-\overline { W}_s+i\epsilon) }  \Bigg]\,,
\end{align}
\end{subequations}
where we introduce 
the notation $\overline {x}=x(\vec q-\vec p)$. 
\begin{figure}
    \centering
    \includegraphics[scale=0.4]{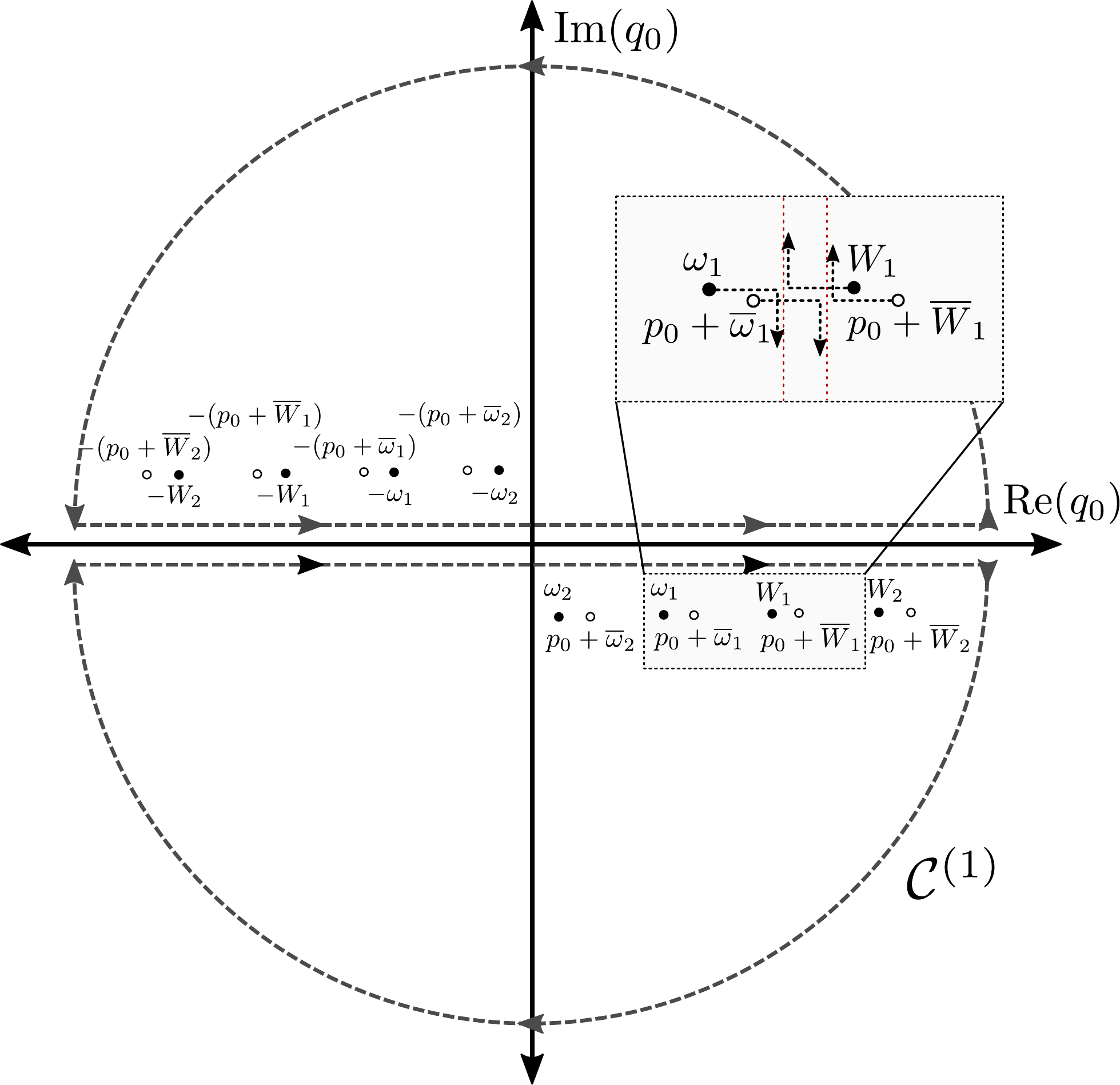}
    \caption{ \label{Fig2} The poles in the lower half-plane are displaced with the 
     $-i\epsilon$ prescription, while those in the upper half-plane with the $i\epsilon$ prescription. 
    According to the explanation as part of the equation~\eqref{pmax} the poles $q_1, q_2$ and $q_5, q_6$ take an imaginary component beyond the momentum value
 $\vert\vec p\vert_{\text{max}}$ and  $\vert \vec q-\vec p\vert_{\text{max}} $, respectively.
 At these values, the poles $q_1, q_5$
move downwards, and the poles  $q_2, q_6$ move upwards parallel to the imaginary axis as indicated in the zoom region. }
\end{figure}
Let us focus on the last integral in~\eqref{Amp2}
\begin{align}\label{Int}
I ^{(1)} _{\mu \nu}&=  \oint _{\mathcal C ^{(1)}}  d q_0
 \text{Tr}\left[\gamma_\nu F(q-p)  \gamma_\mu  F(q)  \right]   \frac{1  }{ D_{q-p}  D_q}   \,.
\end{align}
To compute the integral
we close the contour of integration in the lower half-plane with the curve $\mathcal C^{(1)}$
enclosing the eight poles 
\begin{align}
q_1&=\omega_1-i\epsilon  \,,  \notag \\    q_2&= W_1-i\epsilon   \,,  \notag \\    q_3&=   \omega_2-i\epsilon   
 \,,  \notag \\   
q_4&=W_2-i\epsilon  \,,
\end{align}
 and the displaced ones 
\begin{align}
 q_5&= p_0+\overline \omega_1-i\epsilon  \,,  \notag \\   q_6&=  p_0+\overline W_1-i\epsilon       \,,  \notag \\  
  q_7&= p_0+\overline \omega_2-i\epsilon    \,,  \notag \\  
q_8&= p_0+\overline W_2-i\epsilon \,,
\end{align}
as indicated in the Fig.~\ref{Fig2}.

The integral is 
\begin{align}
I ^{(1)} _{\mu \nu}&= -2\pi i   \sum_{i=1}^8
 \text{Tr}\left[\gamma_\nu F(q-p)  \gamma_\mu  F(q)  \right] _{q_0=q_i} \text{Res}(q_i)   \,,
\end{align}
where we denote $\text{Res}(q_i)$ the residue at $q_i$ of the singular part 
$\frac{1}{D_q D_{q-p}}$. 
A direct calculation gives
\begin{subequations}
\begin{align}
  \text{Res}\left(q_1 \right)  &=\frac{1}{2g_2^4\omega_1
(\omega_1^2-W_1^2) (\omega_1^2-\omega_2^2)(\omega_1^2-W_2^2)}  \notag \\ 
& \phantom{{}={}} \times
\left( \frac{1}{D_{q-p}}\right)_{q_0= q_1} \,,
\end{align}
\begin{align}
  \text{Res}\left(q_2   \right)    &=\frac{1}{2g_2^4W_1 
 (W_1^2-\omega_1^2) (W_1^2-\omega_2^2)(W_1^2-W_2^2)}    \notag \\ 
& \phantom{{}={}} \times
\left( \frac{1}{D_{q-p}}\right)_{q_0= q_2} \,,
\end{align}
\begin{align}
 \text{Res}\left(q_3 \right)
 &=\frac{1}{2g_2^4\omega_2 (\omega_2^2-\omega_1^2)
 (\omega_2^2-W_1^2)(\omega_2^2-W_2^2)}   \notag \\ 
 &\phantom{{}={}}\times
\left( \frac{1}{D_{q-p}}\right)_{q_0= q_3} \,,
\end{align}
\begin{align}
 \text{Res}\left(q_4   \right)
 &=\frac{1}{2g_2^4W_2 (W_2^2-\omega_1^2) (W_2^2-\omega_2^2)(W_2^2-W_1^2)}
 \notag \\ 
 &\phantom{{}={}}\times
\left( \frac{1}{D_{q-p}}\right)_{q_0= q_4} \,,
\end{align}
\end{subequations}
and
\begin{subequations}
\begin{align}
 \text{Res} \left(q_5 \right)&=     \frac{1}{2g_2^4 \overline { \omega}_1
(\overline { \omega}_1^2-\overline { W}_1^2) (\overline
{ \omega}_1^2-\overline { \omega}_2^2)(\overline {
\omega}_1^2-\overline { W}_2^2)}   \notag  \\   &\phantom{{}={}}\times   \left( \frac{1}{D_{q}}\right)_{q_0= q_5 }    \,,
\end{align}
\begin{align}
 \text{Res} \left(q_6  \right) &=  \frac{1}{2g_2^4 \overline { W}_1
(\overline { W}_1^2-\overline { \omega}_1^2) (\overline 
{ W}_1^2-\overline { \omega}_2^2)(\overline { W}_1^2-\overline
{ W}_2^2)}     \notag    \\ 
  &\phantom{{}={}}\times   \left( \frac{1}{D_{q}}\right)_{q_0= q_6 }  \,,
\end{align}
\begin{align}
 \text{Res}\left(q_7  \right)  &=     \frac{1}{2g_2^4\overline 
 { \omega}_2 (\overline {
\omega}_2^2-\overline { \omega}_1^2) (\overline 
{ \omega}_2^2-\overline { W}_1^2)(\overline 
{ \omega}_2^2-\overline { W}_2^2)}    \notag    \\  &\phantom{{}={}}\times   \left( \frac{1}{D_{q}}\right)_{q_0= q_7}    \,,
\end{align}
\begin{align}
 \text{Res}\left(q_8   \right)& =
 \frac{1}{2g_2^4\overline { W}_2 (\overline { W}_2^2-
\overline { \omega}_1^2) (\overline { W}_2^2-\overline
{ \omega}_2^2)(\overline { W}_2^2-\overline { W}_1^2)}   \notag  \\
 &\phantom{{}={}}\times     \left(\frac{1}{D_{q}}\right)_{q_0= q_8}  
\,,
\end{align}
\end{subequations}
where we have eliminated the $i\epsilon$ where it is not relevant.

We consider $p_0$ to be positive and hence 
the last four terms above 
do not contribute to the amplitude's discontinuity or imaginary part.
Decomposing in partial fraction the relevant contributions come from
\begin{subequations}
\begin{align}
  \left( \frac{1}{D_{q-p}}\right)_{q_0= q_1}   = 
 \frac{1}{g_2^4} \prod_{s=1,2}  \frac{\zeta(\omega_1)}{
 (p_0-\omega_1+\overline{ \omega}_s)(p_0-\omega_1+\overline{ W}_s) }  \,,
\end{align}
\begin{align}
\left( \frac{1}{D_{q-p}}\right)_{q_0= q_2}= \frac{1}{g_2^4} \prod_{s=1,2}  \frac{\zeta(W_1)}{
 (p_0-W_1+\overline{ \omega}_s)(p_0-W_1+\overline{ W}_s) } 
  \,,
\end{align}
\begin{align}
\left( \frac{1}{D_{q-p}}\right)_{q_0= q_3} =  
\frac{1}{g_2^4} \prod_{s=1,2}  \frac{\zeta(\omega_2)}{
 (p_0-\omega_2+\overline{ \omega}_s)(p_0-\omega_2+\overline{ W}_s) } \,,
\end{align}
\begin{align}
\left( \frac{1}{D_{q-p}}\right)_{q_0= q_4} =  
\frac{1}{g_2^4} \prod_{s=1,2}  \frac{\zeta(W_2) }{
 (p_0-W_2+\overline{ \omega}_s)(p_0-W_2+\overline{ W}_s) }
  \,,
\end{align}
\end{subequations}
with
\begin{align}
\zeta(x)&=      
\frac{1}{  (\overline{ \omega}_1-\overline{ W}_1) (\overline{ \omega}_1
-\overline{ \omega}_2)  (\overline{ \omega}_1-\overline{ W}_2)(p_0-x
-\overline{ \omega}_1+i\epsilon)} \nonumber \\&   
+\frac{1}{  (\overline{ W}_1-\overline{ \omega}_1) (\overline{ W}_1-\overline{ \omega}_2) 
(\overline{ W}_1-\overline{ W}_2)(p_0-x-\overline{ W}_1+i\epsilon)} \nonumber \\& +\frac{1}{ 
(\overline{ \omega}_2-\overline{ \omega}_1) (\overline{ \omega}_2-\overline{ W}_1) 
(\overline{ \omega}_2-\overline{ W}_2)(p_0-x-\overline{ \omega}_2+i\epsilon)} \nonumber \\
&+\frac{1}{  (\overline{ W}_2-\overline{ \omega}_1) (\overline{ W}_2-\overline{ W}_1) 
(\overline{ W}_2-\overline{ \omega}_2)(p_0-x-\overline{ W}_2+i\epsilon)} \,.
\end{align}
Let us consider the identity
\begin{eqnarray}\label{PVC}
\frac{1}{x\pm i\epsilon}=\mathcal{P}\left(\frac{1}{x}\right)\mp i\pi \delta(x)\,,
\end{eqnarray}
where $\mathcal{P}$ denotes the principal value.
The contributions to the imaginary 
part of the scattering amplitude are
\begin{subequations}
\begin{align} \label{I1}
&  \text{Im}  \left( \frac{1}{D_{q-p}}\right)_{q_0= q_1}   \notag  \\ &\phantom{{}={}}   =  -\frac{\pi}{g_2^4}  \Bigg[  \frac{\delta
(p_0-\omega_1-\overline \omega_1)}{2\overline\omega_1  (\overline \omega_1^2
-\overline W_1^2) (\overline \omega_1^2-\overline \omega_2^2)  
(\overline \omega_1^2-\overline W_2^2)}  \notag  \\ &\phantom{{}={}}    +\frac{\delta(p_0-\omega_1-\overline W_1)}{2\overline W_1  
(\overline W_1^2-\overline \omega_1^2) (\overline W_1^2-\overline \omega_2^2) 
(\overline W_1^2-\overline W_2^2)} \notag  \\
 &\phantom{{}={}}  +\frac{\delta(p_0-\omega_1-\overline \omega_2)}{ 2\overline\omega_2
(\overline \omega_2^2-\overline \omega_1^2) (\overline \omega_2^2-\overline W_1^2) 
(\overline \omega_2^2-\overline W_2^2)}  \notag \\ & \phantom{{}={}} +\frac{\delta(p_0-\omega_1-\overline W_2)}{2\overline W_2^2 
(\overline W_2^2-\overline \omega_1^2) (\overline W_2^2-\overline W_1^2)
(\overline W_2^2-\overline \omega_2^2)} \Bigg]\,, 
\end{align}
\begin{align} \label{I2}
& \text{Im}\left( \frac{1}{D_{q-p}}\right)_{q_0= q_2}     \notag  \\ &\phantom{{}={}}     =- \frac{\pi}{g_2^4} \Bigg[  \frac{\delta(p_0-W_1-\overline
\omega_1)}{2\overline\omega_1  (\overline \omega_1^2-\overline W_1^2) (\overline \omega_1^2-\overline
\omega_2^2)  (\overline \omega_1^2-\overline W_2^2)} \notag   \\  & \phantom{{}={}} +\frac{\delta(p_0-W_1
-\overline W_1)}{2\overline W_1  (\overline W_1^2-\overline \omega_1^2) 
(\overline W_1^2-\overline \omega_2^2)  (\overline W_1^2-\overline W_2^2)} 
\nonumber \\
 &\phantom{{}={}}+\frac{\delta(p_0-W_1-\overline \omega_2)}{2\overline\omega_2  
 (\overline \omega_2^2-\overline
\omega_1^2) (\overline \omega_2^2-\overline W_1^2) 
 (\overline \omega_2^2-\overline W_2^2)}   \notag   \\  & \phantom{{}={}}  +\frac{\delta
 (p_0-W_1-\overline W_2)}{2\overline W_2  (\overline W_2^2-\overline \omega_1^2)
(\overline W_2^2-\overline W_1^2)  (\overline W_2^2-\overline \omega_2^2)} \Bigg]\,, 
\end{align}
\begin{align} \label{I3}
& \text{Im}\left( \frac{1}{D_{q-p}}\right)_{q_0= q_3}   \notag  \\ &\phantom{{}={}}    =  -\frac{\pi}{g_2^4} \Bigg[  \frac{\delta(p_0
-\omega_2-\overline \omega_1)}{2\overline\omega_1  (\overline 
\omega_1^2-\overline W_1^2)
(\overline \omega_1^2-\overline \omega_2^2)  (\overline \omega_1^2
-\overline W_2^2)}  \notag   \\  &  \phantom{{}={}} +\frac{\delta(p_0-\omega_2-\overline W_1)}{2\overline 
W_1  (\overline W_1^2-\overline
\omega_1^2) (\overline W_1^2-\overline \omega_2^2) 
 (\overline W_1^2-\overline W_2^2)} \nonumber \\
 &\phantom{{}={}}+\frac{\delta(p_0-\omega_2-\overline \omega_2)}
 {2\overline\omega_2  (\overline 
\omega_2^2-\overline \omega_1^2) (\overline \omega_2^2-\overline W_1^2) 
 (\overline 
\omega_2^2-\overline W_2^2)}    \notag   \\  &\phantom{{}={}}   +\frac{\delta(p_0-\omega_2
-\overline W_2)}{2\overline W_2  (\overline W_2^2-\overline 
\omega_1^2) (\overline W_2^2-\overline W_1^2)  (\overline 
W_2^2-\overline \omega_2^2)} 
\Bigg]\,, 
\end{align}
\begin{align} \label{I4}
& \text{Im}\left( \frac{1}{D_{q-p}}\right)_{q_0= q_4}  \notag  \\ &\phantom{{}={}}   =  -\frac{\pi}{g_2^4} \Bigg[  
\frac{\delta(p_0-W_2-\overline \omega_1)}{2\overline\omega_1  (\overline \omega_1^2
-\overline W_1^2) (\overline \omega_1^2-\overline \omega_2^2)  (\overline \omega_1^2
-\overline W_2^2)}   \notag   \\  & \phantom{{}={}}  +\frac{\delta(p_0-W_2-\overline W_1)}{2\overline W_1  (\overline W_1^2-\overline
\omega_1^2) (\overline W_1^2-\overline \omega_2^2)  (\overline W_1^2-\overline W_2^2)} 
\nonumber \\
 &\phantom{{}={}}+\frac{\delta(p_0-W_2-\overline \omega_2)}{2\overline\omega_2  (\overline 
\omega_2^2-\overline \omega_1^2) (\overline \omega_2^2-\overline W_1^2)  (\overline
\omega_2^2-\overline W_2^2)}  \notag   \\  &\phantom{{}={}}   +\frac{\delta(p_0-W_2-\overline W_2)}{2\overline W_2  (\overline W_2^2-\overline
\omega_1^2) (\overline W_2^2-\overline W_1^2)  (\overline W_2^2-\overline \omega_2^2)}
\Bigg]\,. 
\end{align}
\end{subequations}

In principle, some contributions 
depend on deltas 
involving ghost modes, as seen in~\eqref{I1}-\eqref{I4}.
However, these contributions demand an energy of the order $1/g_2$,
which lie far beyond the region of validity of the effective theory.
Hence, the deltas involving a $W_{1,2}$ or $\overline W_{1,2}$ mode 
vanish, in other words, the initial $p_0=\omega_s+\omega_r$ cannot
balance the energetic restriction given by these deltas so we disregard them. 
In this way, we are left with the contributions
\begin{widetext}
\begin{align}
&2\text{Im}(\mathcal{M}_F^{(1)})=\frac{-e^4}{p^4} J_1^\mu(p_1,p_2) J_2^\nu(p_1,p_2) 
 \int \frac{d^3 \vec q}{(2\pi)^4} (2\pi)^2    \\
 &\times      \Bigg[\frac{\text{Tr}[\gamma_\nu F(\omega_1-p_0,\vec q-\vec p)
 \gamma_\mu F(\omega_1,\vec q)]}{2g_2^4\omega_1 (\omega_1^2-W_1^2)
  (\omega_1^2-\omega_2^2)(\omega_1^2-W_2^2)}      \left( \frac{\delta
  (p_0-\omega_1-\overline \omega_1)}{2g_2^4\overline\omega_1
    (\overline \omega_1^2-\overline W_1^2) (\overline \omega_1^2
    -\overline \omega_2^2)  (\overline \omega_1^2-\overline W_2^2)}
    +\frac{\delta(p_0-\omega_1-\overline \omega_2)}{ 2g_2^4\overline
    \omega_2 (\overline \omega_2^2-\overline \omega_1^2) (\overline
     \omega_2^2-\overline W_1^2)  (\overline \omega_2^2-\overline W_2^2)}  \right) \notag \\ \notag
 &  +\frac{\text{Tr}[\gamma_\nu F(\omega_2-p_0,\vec q-\vec p)\gamma_\mu F
 (\omega_2,\vec q)]}{2g_2^4\omega_2 (\omega_2^2-\omega_1^2) (\omega_2^2
 -W_1^2)(\omega_2^2-W_2^2)}      \left(  \frac{\delta(p_0-\omega_2
 -\overline \omega_1)}{2g_2^4\overline\omega_1  (\overline \omega_1^2
 -\overline W_1^2) (\overline \omega_1^2-\overline \omega_2^2)  (\overline
  \omega_1^2-\overline W_2^2)} +\frac{\delta(p_0-\omega_2-\overline \omega_2)}
  {2g_2^4\overline\omega_2  (\overline \omega_2^2-\overline \omega_1^2)
   (\overline \omega_2^2-\overline W_1^2)  (\overline \omega_2^2-\overline W_2^2)} \right)  \Bigg]\,.
\end{align}

We introduce the variables $k_1^0,k_2^0$ followed by delta 
functions as follows
\begin{align}
2&\text{Im}(\mathcal{M}_F^{(1)})=\frac{-e^4}{p^4} J_1^\mu(p_1,p_2) 
J_2^\nu(p_1,p_2)     \int \frac{d^3 q}{(2\pi)^4} (2\pi)^2 \int dk_1^0
\int dk_2^0 \delta (p_0-k_1^0-k_2^0)   \Bigg[\frac{\text{Tr}[\gamma_\nu 
 F(-\overline\omega_1,\vec q-\vec p)\gamma_\mu F(\omega_1,\vec q)]}
 {2g_2^4\omega_1 (\omega_1^2-W_1^2) (\omega_1^2-\omega_2^2)(\omega_1^2-W_2^2)}  
  \notag  \\       & \times   \frac{\delta(k_1^0-\omega_1)\delta(k_2^0-
 \overline{\omega} _1  )}{2g_2^4\overline\omega_1  (\overline \omega_1^2-\overline W_1^2)
 (\overline \omega_1^2-\overline \omega_2^2)  (\overline \omega_1^2-\overline W_2^2)}
 +\frac{\text{Tr}[\gamma_\nu F(-\overline\omega_2,\vec q-\vec p)
 \gamma_\mu F(\omega_1,\vec q)]}{2g_2^4\omega_1 (\omega_1^2-W_1^2)
 (\omega_1^2-\omega_2^2)(\omega_1^2-W_2^2)}  
 \frac{\delta(k_1^0-\omega_1)\delta(k_2^0-\overline{\omega}_2 )}
 { 2g_2^4\overline\omega_2 (\overline \omega_2^2-\overline \omega_1^2) (\overline
 \omega_2^2-\overline W_1^2)  (\overline \omega_2^2-\overline W_2^2)} \nonumber \\
 &+\frac{\text{Tr}[\gamma_\nu F(-\overline\omega_1,\vec q-\vec p
 \gamma_\mu F(\omega_2,\vec q)]}{2g_2^4\omega_2 (\omega_2^2
 -\omega_1^2) (\omega_2^2-W_1^2)(\omega_2^2-W_2^2)}  
 \frac{\delta(k_1^0-\omega_2)\delta(k_2^0-\overline{
 \omega_1})}{2g_2^4\overline\omega_1  (\overline \omega_1^2-\overline W_1^2)
 (\overline \omega_1^2-\overline \omega_2^2)  (\overline \omega_1^2-\overline W_2^2)}
 +\frac{\text{Tr}[\gamma_\nu F(-\overline\omega_2,\vec 
 q-\vec p)\gamma_\mu F(\omega_2,\vec q)]}{2g_2^4\omega_2
 (\omega_2^2-\omega_1^2) (\omega_2^2-W_1^2)(
 \omega_2^2-W_2^2)}      \notag  \\       & \times     
 \frac{\delta(k_1^0-\omega_2) 
 \delta(k_2^0-\overline{\omega}_2)}{2g_2^4\overline\omega_2 
 (\overline \omega_2^2-\overline \omega_1^2) (\overline \omega_2^2-\overline W_1^2)
 (\overline \omega_2^2-\overline W_2^2)} \Bigg] \notag \,,
\end{align}
and then by using
\begin{align}\label{deltaid}
  \int \frac{d^3q}{(2\pi)^3}&=\int \frac{d^3k_1}{(2\pi)^3}
  \int\frac{d^3k_2}{(2\pi)^3}(2\pi)^3\delta^{(3)}(\vec p-\vec k_1-\vec k_2)\,,
\end{align}
and defining $\vec k_1=\vec q$, $\vec k_2=\vec p-\vec q$ we write
\begin{align}
2&\text{Im}(\mathcal{M}_F^{(1)})=\frac{-e^4}{p^4} J_1^\mu(p_1,p_2) J_2^\nu(p_1,p_2) 
\int \frac{d^4 k_1}{(2\pi)^4} \int 
\frac{d^4 k_2}{(2\pi)^4} (2\pi)^4  \delta^{(4)} (p-k_1-k_2)\Bigg[\frac{\text{Tr}
[\gamma_\nu F(-k_2^0,-\vec k_2)
 \gamma_\mu F(k_1^0,\vec k_1)]}{2g_2^4\omega_1 (\omega_1^2-W_1^2)
 (\omega_1^2-\omega_2^2)(\omega_1^2-W_2^2)}     \notag \\ &\phantom{{}={}} \times
 \frac{(2\pi)^2\delta(k_1^0-\omega_1)\delta(k_2^0-\overline{\omega}_1)}
 {2g_2^4\overline\omega_1  (\overline \omega_1^2-\overline W_1^2) (\overline 
 \omega_1^2-\overline {\omega}_2^2)  (\overline \omega_1^2-\overline W_2^2)} 
 +\frac{\text{Tr}[\gamma_\nu F(-k_2^0,-\vec k_2)\gamma_\mu F
 (k_1^0,\vec k_1)]}{2g_2^4\omega_1 (\omega_1^2-W_1^2) (\omega_1^2-\omega_2^2)
 (\omega_1^2-W_2^2)} 
 \frac{(2\pi)^2\delta(k_1^0-\omega_1)\delta(k_2^0-\overline{\omega}_2)}
 { 2g_2^4\overline\omega_2 (\overline \omega_2^2-\overline \omega_1^2) (\overline \omega_2^2-\overline W_1^2) 
 (\overline \omega_2^2-\overline W_2^2)} \notag \\ & +\frac{\text{Tr}[\gamma_\nu F(-k_2^0,-\vec k_2)
 \gamma_\mu F(k_1^0,\vec k_1)]
 }{2g_2^4\omega_2 (\omega_2^2-\omega_1^2) (\omega_2^2-W_1^2)(\omega_2^2-W_2^2)} 
 \frac{(2\pi)^2\delta(k_1^0-\omega_2)\delta(k_2^0-\overline{\omega}_1)}{2g_2^4\overline\omega_1 
 (\overline \omega_1^2-\overline W_1^2) (\overline \omega_1^2-\overline \omega_2^2) 
 (\overline \omega_1^2-\overline W_2^2)}+\frac{\text{Tr}[\gamma_\nu F(-k_2^0,-\vec k_2)\gamma_\mu F
 (k_1^0,\vec k_1)]}{2g_2^4\omega_2 (\omega_2^2-\omega_1^2)
 (\omega_2^2-W_1^2)(\omega_2^2-W_2^2)}   \notag \\ &\phantom{{}={}} \times
 \frac{(2\pi)^2\delta(k_1^0-\omega_2)\delta(k_2^0-
 \overline{\omega}_2)}{2g_2^4\overline\omega_2  (\overline \omega_2^2-\overline \omega_1^2)
 (\overline \omega_2^2-\overline W_1^2)  (\overline {\omega}_2^2-\overline W_2^2)} \Bigg]  \notag \,.
\end{align}
 Now, we will relate the amplitude with the total cross-section of 
 the cutting diagram. To archive this connection, we 
 recall the relations~\eqref{u1u1}, \eqref{u2u2}, \eqref{v1v1}, \eqref{v1v1} and arrive at
\begin{align}
2&\text{Im}(\mathcal{M}_F^{(1)})=\frac{e^4}{p^4} J_1^\mu(p_1,p_2) J_2^\nu(p_1,p_2)  
\int \frac{d^4 k_1}{(2\pi)^4} \int \frac{d^4 k_2}{(2\pi)^4} (2\pi)^4  \delta^{(4)} 
(p-k_1-k_2) \Bigg[\frac{\text{Tr}[\gamma_\nu v^{(1)}(k_2)\overline{v}^{(1)}(k_2)
\gamma_\mu u^{(1)}(k_1)\overline{u}^{(1)}(k_1)]}{2g_2^2\omega_1 (\omega_1^2-W_1^2) }  
 \notag \\ & \times
 \frac{(2\pi)^2\delta(k_1^0-\omega_1)\delta(k_2^0-\overline{\omega}_1)}{2g_2^2
 \overline\omega_1  (\overline \omega_1^2-\overline W_1^2)   }
 +\frac{\text{Tr}[\gamma_\nu v^{(2)}(k_2)\overline{v}^{(2)}(k_2)\gamma_\mu u^{(1)}
 (k_1)\overline{u}^{(1)}(k_1)]}{2g_2^2\omega_1 (\omega_1^2-W_1^2) }
  \frac{(2\pi)^2\delta(k_1^0-\omega_1)\delta(k_2^0-\overline{\omega}_2)}
  {2g_2^2\overline\omega_2    (\overline \omega_2^2-\overline W_2^2)} \nonumber \\
 &+\frac{\text{Tr}[\gamma_\nu v^{(1)}(k_2)\overline{v}^{(1)}(k_2)
 \gamma_\mu u^{(2)}(k_1)\overline{u}^{(2)}(k_1)]}{2g_2^2\omega_2 (\omega_2^2-W_2^2)}  
 \frac{(2\pi)^2\delta(k_1^0-\omega_2)\delta(k_2^0-\overline{\omega}_1)}
 {2g_2^2\overline\omega_1  (\overline \omega_1^2-\overline W_1^2)   } 
 +\frac{\text{Tr}[\gamma_\nu v^{(2)}(k_2)\overline{v}^{(2)}(k_2)
 \gamma_\mu u^{(2)}(k_1)\overline{u}^{(2)}(k_1)]}{2g_2^2\omega_2 (\omega_2^2-W_2^2)} 
  \notag \\ & \times
  \frac{(2\pi)^2\delta(k_1^0-\omega_2)\delta(k_2^0-\overline{\omega}_2)}
  {2g_2^2\overline\omega_2    (\overline \omega_2^2-\overline W_2^2)} \Bigg]\,.
\end{align}
\end{widetext}
At this point, it is convenient to define a physical delta where we will exclude the ghost frequencies
\begin{align}
\delta^{(\text{phys)}}(\Lambda_s^2(p_0))=\sum_{\text{phys,a}}
\frac{\delta(p_0-p_a)}{\vert (\Lambda_s^2)'(p_a)\vert} \,,
\end{align}
for $s=1,2$ with the new notation
$\Lambda_1^2\equiv \Lambda_+^2$ and $\Lambda_2^2\equiv
 \Lambda_-^2$ and
where $p_a$ are the zeros of the function $\Lambda_s^2(p_0)$.
In our case, we have
\begin{eqnarray}
\delta^{(\text{phys)}}  {(\Lambda_s^2(p_0))}&=&\frac{\delta(p_0-\omega_s)
-\delta(p_0+\omega_s)}{ 2\omega_sg_2^2(W_s^2-\omega_s^2)}  \,.
\end{eqnarray}

This allows us to write the LHS as of the cutting equation as
\begin{align}
2&\text{Im}(\mathcal{M}_F^{(1)})=\frac{e^4}{p^4} J_1^\mu(p_1,p_2) 
J_2^\nu(p_1,p_2)  \\ & \times
\int \frac{d^4 k_1}{(2\pi)^4} \int \frac{d^4 k_2}{(2\pi)^4} (2\pi)^4 
 \delta^{(4)} (p-k_1-k_2)\nonumber \\
&\times\sum_{\overline{s},\overline{r}=1,2}\Bigg[\text{Tr}[
\gamma_\nu v^{\overline{r}}(k_2)\overline{v}^{\overline{r}}(k_2)
\gamma_\mu u^{\overline{s}}(k_1)\overline{u}^{\overline{s}}(k_1)] \notag \\ & \times
 (2\pi)^2\delta ^{(\text{phys)}}  \left(\Lambda_{\overline{s}}^2   \left(k_1^0\right)   \right) \delta ^{(\text{phys)}} 
 \left({\Lambda}_{\overline{r}}^2 \left(k_2^0  \right)  \right)\theta \left(k_1^0 \right)\theta \left(k_2^0 \right) \Bigg]\,,\notag
\end{align}
and 
\begin{align}
2&\text{Im}(\mathcal{M}_F^{(1)})
=\frac{e^4}{p^4} J_1^\mu(p_1,p_2) J_2^\nu(p_1,p_2) \\ 
& \times \int \frac{d^4 k_1}{(2\pi)^4} \int \frac{d^4 k_2}{(2\pi)^4} (2\pi)^4  \delta^{(4)} 
(p-k_1-k_2)\nonumber \\
&\times\sum_{\overline{s},\overline{r}=1,2}\Bigg[\overline{u}^{\overline{s}}(k_1)
\gamma_\nu v^{\overline{r}}(k_2)\overline{v}^{\overline{r}}(k_2)\gamma_\mu 
u^{\overline{s}}(k_1)] \notag  \\ & \times
 (2\pi)^2\delta ^{(\text{phys)}}  \left(\Lambda_{\overline{s}}^2  \left(k_1^0 \right)   \right)\delta^{(\text{phys)}}   \left({\Lambda}_
 {\overline{r}}^2  \left(k_2^0  \right)  \right)\theta \left(k_1^0\right)\theta \left(k_2^0 \right) \Bigg]\,.\notag
\end{align}
Finally, we can recognize the scattering amplitude for the cut diagram using the symmetry of the $(\mu,\nu)$ indeces  
\begin{align}
i\widehat{\mathcal{M}  ^{(1)}}&:=\bigg[\overline{v}^r(p_2)(-ie\gamma^\mu)u^s(p_1)
\bigg]\left(\frac{-i\eta_{\mu\nu}}{p^2+i\epsilon}\right)\nonumber \\
&\times \bigg[\overline{u}^{\overline{s}}(k_1)(-ie\gamma^\nu)v^{\overline{r}}(-k_2) \bigg]\,.
\end{align}
This proves the optical theorem and the unitary
evolution of the $S$-matrix
for the one-loop Bhabha scattering diagram
\begin{align}
&2\text{Im}(\mathcal{M}_F^{(1)})=\sum_{\overline{s},\overline{r}=1,2}\int \frac
{d^4k_1}{(2\pi)^4}\frac{d^4 k_2}{(2\pi)^4}(2\pi)^4 \delta^{(4)}(p-k_1-k_2)\nonumber \\
& \times  \vert\widehat{\mathcal{M} ^{(1)}}\vert^2  (2\pi)^2\delta ^{(\text{phys)}}  \left(\Lambda_{\overline{s}}^2
\left(k_1^0 \right) \right)\delta ^{(\text{phys)}}   \left({\Lambda}_{\overline{r}}^2\left (k_2^0 \right)  \right)  \notag  \\&\times  \theta 
 \left(k_1^0 \right)\theta  \left(k_2^0\right)\,.
\end{align}
\subsection{One-loop Compton scattering}\label{subsec:IV-2}
\begin{figure}
    \centering
    \includegraphics[scale=0.5]{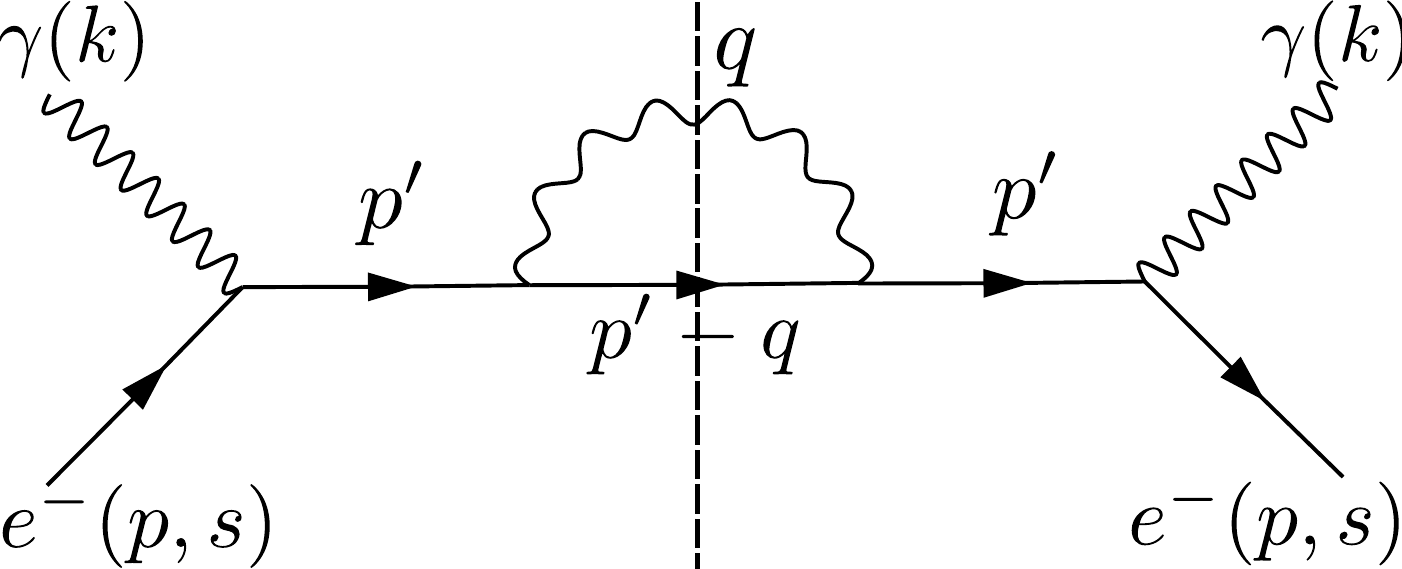}
    \caption{ \label{Fig3} The diagram representing the scattering 
    process $\gamma e^-\to\gamma e^-$ and the cut diagram produced by the 
     vertical segmented line. }
\end{figure}

We continue with the one-loop Compton
scattering process $e^-(p,s)+\gamma(k)\rightarrow e^{-}(p,s)+\gamma(k)$ of
Fig.~\ref{Fig3}.

\begin{figure}
    \centering
    \includegraphics[scale=0.40]{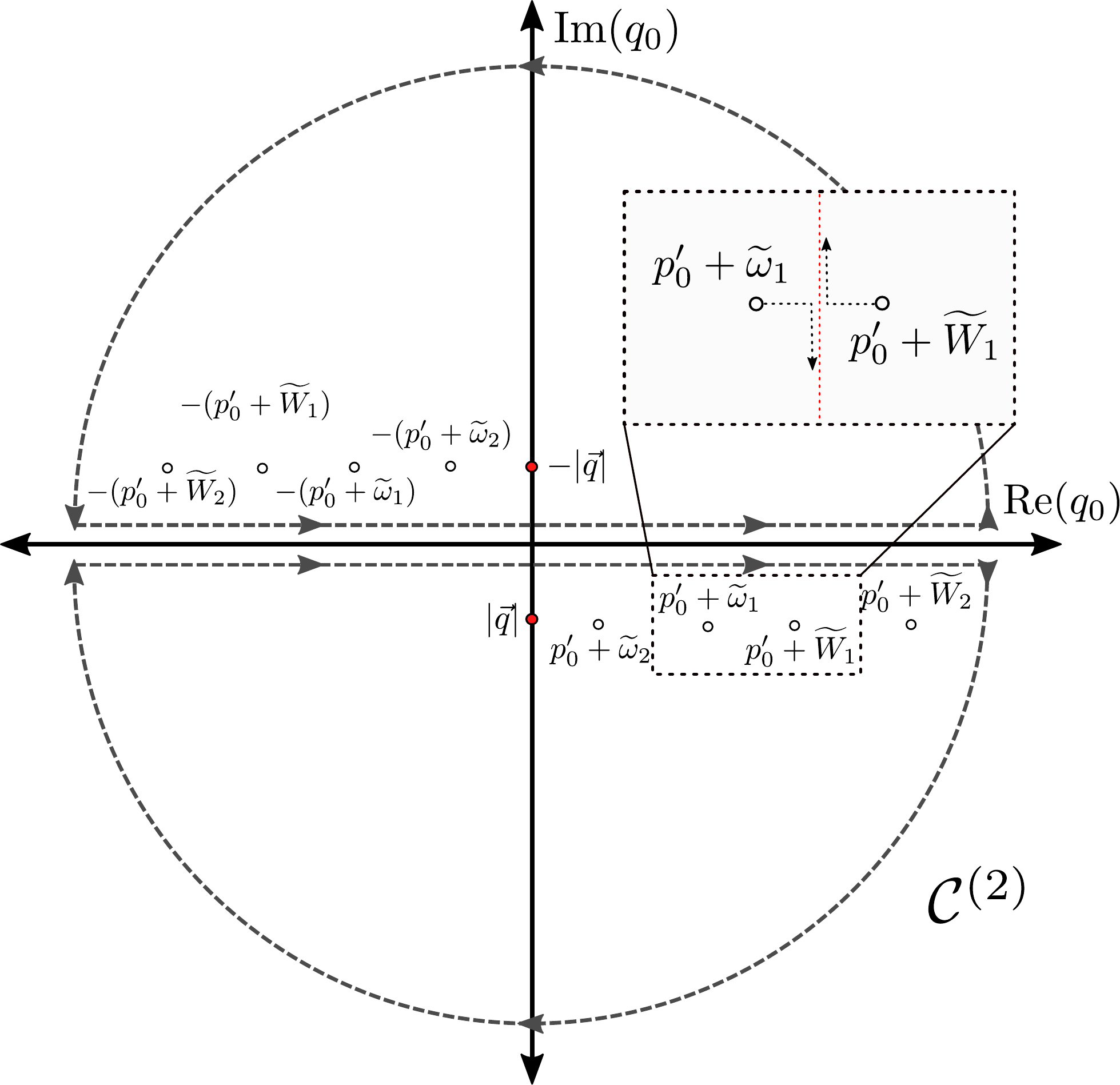}
    \caption{ \label{Fig4} The contour $\mathcal C^{(2)}$ encloses the poles $q_1, q_2, q_3, q_4, q_5$ and at the critical energy the two poles 
    $q_2$ and $q_4$ become complex, as indicated in the figure.}
\end{figure}

We have the amplitude 
\begin{align}
i\mathcal{M}_F^{(2)}=&(-1)\left[\varepsilon_\rho^{*}(k)\bar{u}^s(p)
(-ie\gamma^\rho) \right]\left(\frac{iF(p')}{D_{p'}}\right)\nonumber \\
\times&\int \frac{d^4 q}{(2\pi)^4}\left[(-ie\gamma^\mu)\frac{iF(p'-q)}
{D_{p'-q}}(-ie\gamma^\nu) \frac{-i\eta_{\mu\nu}}{q^2+i\epsilon} \right]\nonumber \\
&\times\left(\frac{iF(p')}{D_{p'}}\right)\left[ (-ie\gamma^\sigma)
u^s(p)\varepsilon_\sigma(k) \right] \,.
\end{align} 
By defining the quantities
\begin{eqnarray}
    J(p,k)&:=&\frac{ F(p')\gamma^\sigma u^s(p)\varepsilon_\sigma(k) }{D_{p'}}\,,  \\
    J^{*}(p,k)&:=&\frac{\varepsilon_\rho^{*}(k)\bar{u}^s(p)\gamma^\rho F(p')}{D_{p'}} \,,
\end{eqnarray}
we rewrite as
\begin{align}
    \mathcal{M}_F^{(2)}&=-ie^4 J^*(p,k) \int    \frac{d^3 \vec q}{(2\pi)^4}\notag  \\ & \phantom{{}={}}\times  \int dq_0\frac{\gamma^\mu
     F(p'-q)\gamma_\mu}{(q^2+i\epsilon)D_{p'-q}} J(p,k)\,.
\end{align}
Consider the last integral
\begin{align}
   I^{(2)}= \int_{\mathcal C}  d q_0    \frac{
     F(p'-q) }{(q^2+i\epsilon)D_{p'-q}} \,,
\end{align}
where for the singular part of the fermion propagator $\frac{1}{D_{p'-q}} $ 
we have eight poles 
and for 
the photon part $\frac{1}{q^2+i\epsilon}$ we have two more, see Fig.~\ref{Fig4}. 

To compute the integral, we employ
 Cauchy residue theorem 
and consider the contour of integration $\mathcal C^{(2)}$ that closes from below, as shown in
Fig.~\ref{Fig4}.
The contour encloses five poles, and we obtain 
\begin{align}
   I^{(2)}= -2\pi i \sum _{i=1}^5  [F(p'-q)]_{q_0=q_i}\text{Res}(q_i)   \,,
\end{align}
with 
\begin{align}
q_1&=\vert\vec{q}\vert   -i\epsilon  \,,   \\q_2&=p_0'+\widetilde{\omega}_1 -i\epsilon  \,,  
\\ q_3&=  p_0'+\widetilde{\omega}_2-i\epsilon  \,, \\q_4&=p_0'+\widetilde{W}_1-i\epsilon  \,,   \\
q_5&=p_0'+\widetilde{W}_2-i\epsilon \,,
\end{align}
with the residues of the singular part $\frac{1}{(q^2+i\epsilon)D_{p'-q}}$ at $x$
denoted 
by $\text{Res}(x)$.

A calculation gives
\begin{align}
& \text{Res}(q_1)      =\frac{1}{2g_2^4\vert\vec{q}\vert(p'_0-\vert\vec{q}\vert
+\widetilde{\omega}_1)(p'_0-\vert\vec{q}\vert+\widetilde{W}_1)  }  \notag  \\   &\times \frac{1}{
(p'_0-\vert\vec{q}\vert+\widetilde{\omega}_2)(p'_0-\vert\vec{q}\vert+\widetilde{W}_2)  
 (p'_0-\vert\vec{q}\vert-\widetilde{\omega}_1+2i\epsilon)} \notag\\  &\times 
\frac{1}{         (p'_0-\vert\vec{q}\vert-\widetilde{W}_1
+2i\epsilon)(p'_0-\vert\vec{q}\vert-\widetilde{\omega}_2+2i\epsilon)}  \notag \\ &\times \frac{1}{ 
 (p'_0-\vert\vec{q}\vert-\widetilde{W}_2+2i\epsilon)}\,,
\end{align}
\begin{align}
&\text{Res}(q_2 )=\frac{1}{2g_2^4\widetilde{\omega}_1(\widetilde{W}_1^2-\widetilde{\omega}_1^2)
(\widetilde{\omega}_2^2-\widetilde{\omega}_1^2)(\widetilde{W}_2^2-\widetilde{\omega}_1^2)   } \notag \\ &\times
  \frac{1}{ (p'_0+\widetilde{\omega}_1-\vert\vec{q}\vert)(p'_0+\widetilde{\omega}_1+\vert\vec{q}\vert-2i\epsilon)}\,,
 \end{align}
 \begin{align}
   &  \text{Res}(q_3 )  =  \frac{1}{2g_2^4\widetilde{\omega}_2(\widetilde{\omega}_1^2-\widetilde{\omega}_2^2)(\widetilde{W}_1^2
   -\widetilde{\omega}_2^2)(\widetilde{W}_2^2-\widetilde{\omega}_2^2)}\notag  \\ &\times \frac{1}{(p'_0
   +\widetilde{\omega}_2-\vert\vec{q}\vert)(p'_0+\widetilde{\omega}_2+\vert\vec{q}\vert-2i\epsilon)}\,,
 \end{align}
 \begin{align}
     & \text{Res}(q_4) =\frac{1}{2g_2^4\widetilde{W}_1(\widetilde{\omega}_1^2-\widetilde{W}_1^2)
     (\widetilde{\omega}_2^2-\widetilde{W}_1^2)(\widetilde{W}_2^2-\widetilde{W}_1^2)}  \notag  \\  
      &\times  \frac{1}{(p'_0+\widetilde{W}_1-\vert\vec{q}\vert)(p'_0+\widetilde{W}_1+\vert\vec{q}\vert-2i\epsilon)}\,,
 \end{align}
 \begin{align}
      &   \text{Res}(q_5 ) =\frac{1}{2g_2^4\widetilde{W}_2(\widetilde{\omega}_1^2
      -\widetilde{W}_2^2)(\widetilde{W}_1^2-\widetilde{W}_2^2)(\widetilde{\omega}_2^2-\widetilde{W}_2^2)}   \notag  \\  &\times 
      \frac{1}{(p'_0
      +\widetilde{W}_2-\vert\vec{q}\vert)(p'_0+\widetilde{W}_2+\vert\vec{q}\vert-2i\epsilon)}      \,.
\end{align}
Considering that $p'_0\geq 0$ only $\text{Res}(q_1)$ has poles that 
contribute to the discontinuity. We use the partial fraction decomposition for the relevant poles obtaining
\begin{align}
& \text{Res}(q_1) =\frac{1}{2g_2^4\vert\vec{q}\vert}\prod_{i=1,2}\frac{1}{(p'_0-\vert\vec{q}\vert+\widetilde{\omega}_i)
(p'_0-\vert\vec{q}\vert+\widetilde{W}_i)} \notag   \\
&\times\left[\frac{1}{(\widetilde{\omega}_1-\widetilde{\omega}_2)(\widetilde{\omega}_1
-\widetilde{W}_1)(\widetilde{\omega}_1-\widetilde{W}_2)(p'_0-\vert\vec{q}\vert-\widetilde{\omega}_1+2i\epsilon)}   \right. 
\notag   \\
&-\frac{1}{(\widetilde{\omega}_1-\widetilde{\omega}_2)(\widetilde{\omega}_2-\widetilde{W}_1)
(\widetilde{\omega}_2-\widetilde{W}_2)(p'_0-\vert\vec{q}\vert-\widetilde{\omega}_2+2i\epsilon)} \notag \\
&+\frac{1}{(\widetilde{\omega}_1-\widetilde{W}_1)(\widetilde{\omega}_2-\widetilde{W}_1)
(\widetilde{W}_1-\widetilde{W}_2)(p'_0-\vert\vec{q}\vert-\widetilde{W}_1+2i\epsilon)} \notag  \\
&\left.-\frac{1}{(\widetilde{\omega}_1-\widetilde{W}_2)(\widetilde{\omega}_2-\widetilde{W}_2)
(\widetilde{W}_1-\widetilde{W}_2)(p'_0-\vert\vec{q}\vert-\widetilde{W}_2+2i\epsilon)} \right]\,. 
\end{align}
Considering the expresion~\eqref{PVC} we have
\begin{align}
 \text{Im } & \left( \text{Res}(q_1)  \right)=\frac{-\pi }{2g_2^4\vert\vec{q}\vert}\left[\frac{
 \delta(p'_0-\vert\vec{q}\vert-\widetilde{\omega}_1)}{2\widetilde{\omega}_1(\widetilde{\omega}_1^2
 -\widetilde{\omega}_2^2)(\widetilde{\omega}_1^2-\widetilde{W}_1^2)(\widetilde{\omega}_1^2
 -\widetilde{W}_2^2)}   \right.   \notag  \\
&-\frac{\delta(p'_0-\vert\vec{q}\vert-\widetilde{\omega}_2)}{2\widetilde{\omega}_2(\widetilde{\omega}_1^2
-\widetilde{\omega}_2^2)(\widetilde{\omega}_2^2-\widetilde{W}_1^2)
(\widetilde{\omega}_2^2-\widetilde{W}_2^2)} \notag \\
&+\frac{\delta(p'_0-\vert\vec{q}\vert-\widetilde{W}_1)}{2\widetilde{W}_1(\widetilde{\omega}_1^2
-\widetilde{W}_1^2)(\widetilde{\omega}_2^2-\widetilde{W}_1^2)(\widetilde{W}_1^2-\widetilde{W}_2^2)} \notag
 \\
&\left.-\frac{\delta(p'_0-\vert\vec{q}\vert-\widetilde{W}_2)}{2\widetilde{W}_2(\widetilde{\omega}_1^2
-\widetilde{W}_2^2)(\widetilde{\omega}_2^2-\widetilde{W}_2^2)(\widetilde{W}_1^2-\widetilde{W}_2^2)}\right]\,.
\end{align}
We apply effective theory again to consider the possible contributions that involve intermediate states
of ghost modes. This lead us to write the left
 hand side of the optical theorem as follows
\begin{align}
    2\text{Im}& (\mathcal{M}_F^{(2)})= e^4 [J^*(p,k)]\int \frac{d^3q}{(2\pi)^4}(2\pi)^2
    \frac{1}{2\vert\vec{q}\vert}  \\
    &\times\left[\frac{\gamma^\mu F(\widetilde{\omega}_1,\vec{p'}-\vec{q})\gamma_\mu
     \delta(p'_0-\vert\vec{q}\vert-\widetilde{\omega}_1)}{2g_2^4\widetilde{\omega}_1
     (\widetilde{\omega}_1^2-\widetilde{\omega}_2^2)(\widetilde{\omega}_1^2
     -\widetilde{W}_1^2)(\widetilde{\omega}_1^2-\widetilde{W}_2^2)}   \right. \nonumber  \\
&\left.-\frac{\gamma^\mu F(\widetilde{\omega}_2,\vec{p'}-\vec{q})\gamma_\mu
 \delta(p'_0-\vert\vec{q}\vert-\widetilde{\omega}_2)}{2g_2^4\widetilde{\omega}_2
 (\widetilde{\omega}_1^2-\widetilde{\omega}_2^2)(\widetilde{\omega}_2^2-
 \widetilde{W}_1^2)(\widetilde{\omega}_2^2-\widetilde{W}_2^2)} \right][J(p,k)]   \,. \notag
\end{align}
We introduce the variables $k_1^0$ and $k_2^0$ followed by they delta functions as follow
\begin{align}
    &2\text{Im}(\mathcal{M}_F^{(2)})= e^4 [J^*(p,k)]\int \frac{d^3q}{(2\pi)^4}(2\pi)^2\frac{1}{2\vert\vec{q}\vert}  \\
    &\times \int dk_1^0 \int dk_2^0 \delta(p_0'-k_1^0-k_2^0) \nonumber \\
    &\times\left[\frac{\gamma^\mu F(\widetilde{\omega}_1,\vec{p'}-\vec{q})\gamma_\mu 
    \delta(k_1^0-\vert\vec{q}\vert)\delta(k_2^0-\widetilde{\omega}_1)}{2g_2^4\widetilde{\omega}_1
    (\widetilde{\omega}_1^2-\widetilde{\omega}_2^2)(\widetilde{\omega}_1^2-\widetilde{W}_1^2)
    (\widetilde{\omega}_1^2-\widetilde{W}_2^2)}   \right. \nonumber  \\
&\left.-\frac{\gamma^\mu F(\widetilde{\omega}_2,\vec{p'}-\vec{q})\gamma_\mu 
\delta(k_1^0-\vert\vec{q}\vert)\delta(k_2^0-\widetilde{\omega}_2)}{2g_2^4\widetilde{\omega}_2
(\widetilde{\omega}_1^2-\widetilde{\omega}_2^2)(\widetilde{\omega}_2^2-\widetilde{W}_1^2
)(\widetilde{\omega}_2^2-\widetilde{W}_2^2)}\right][J(p,k)]  \,, \notag
\end{align}
and then by using the same previous identity~\eqref{deltaid} and 
defining $\vec k_1=\vec q,  \vec k_2=\vec p'-\vec q$ we write
\begin{align}
    2\text{Im}&(\mathcal{M}_F^{(2)})= e^4 [J^*(p,k)]    [J(p,k)]   \\ &\times    \int \frac{d^4k_1}{(2\pi)^4} 
   \int \frac{d^4k_2}{(2\pi)^4}\frac{(2\pi)^4}{2\vert\vec{k}_1\vert} \delta^{(4)}(p'-k_1-k_2) \nonumber \\
    &\times\left[\frac{\gamma^\mu F(k_2^0,\vec{k}_2)\gamma_\mu (2\pi)^2
    \delta(k_1^0-\vert\vec{k}_1\vert)\delta(k_2^0-\widetilde{\omega}_1)}{2g_2^4\widetilde{\omega}_1
    (\widetilde{\omega}_1^2-\widetilde{\omega}_2^2)(\widetilde{\omega}_1^2-\widetilde{W}_1^2)
    (\widetilde{\omega}_1^2-\widetilde{W}_2^2)}   \right. \nonumber  \\
&\left.-\frac{\gamma^\mu F(k_2^0,\vec{k}_2)\gamma_\mu (2\pi)^2
\delta(k_1^0-\vert\vec{k}_1\vert)\delta(k_2^0-\widetilde{\omega}_2)}{2g_2^4\widetilde{\omega}_2
(\widetilde{\omega}_1^2-\widetilde{\omega}_2^2)(\widetilde{\omega}_2^2-\widetilde{W}_1^2)
(\widetilde{\omega}_2^2-\widetilde{W}_2^2)}\right] \notag \,.
\end{align}
Now using the identities \eqref{u1u1} \eqref{u2u2} we find
\begin{align}
    &2\text{Im}(\mathcal{M}_F^{(2)})= e^4 [J^*(p,k)]     [J(p,k)]   \\ 
       & \int \frac{d^4k_1}{(2\pi)^4}   \int \frac{d^4k_2}{(2\pi)^4}\frac{(2\pi)^4
       \delta^{(4)}(p'-k_1-k_2)}{2\vert\vec{k}_1\vert}  \nonumber \\
    &\times\left[\frac{\gamma^\mu u^{(1)}(k_2)\bar{u}^{(1)}(k_2)\gamma_\mu (2\pi)^2
    \delta(k_1^0-\vert\vec{k}_1\vert)\delta(k_2^0-\widetilde{\omega}_1)}{2g_2^2
    \widetilde{\omega}_1(\widetilde{W}_1^2-\widetilde{\omega}_1^2)}   \right. \nonumber  \\
&\left.\frac{\gamma^\mu u^{(2)}(k_2)\bar{u}^{(2)}(k_2)\gamma_\mu (2\pi)^2
\delta(k_1^0-\vert\vec{k}_1\vert)
\delta(k_2^0-\widetilde{\omega}_2)}{2g_2^2\widetilde{\omega}_2(\widetilde{W}_2^2
-\widetilde{\omega}_2^2)}\right] \,.\notag
\end{align}
Let us recall the physical delta definition and use the fact that
\begin{align}
\delta(k_1^2)\theta(k_1^0)=\frac{\delta(k_1^0-\vert\vec{k}_1\vert)}{2\vert\vec{k}_1\vert}\,,
\end{align}
and the identity
\begin{equation}
\sum_{\text{Pol.}}\varepsilon_\mu(p)\varepsilon_\nu^*(p)=\eta_{\mu\nu}\,.
\end{equation}
We obtain
\begin{align}
   & 2\text{Im}(\mathcal{M}_F^{(2)}  )= e^4 [J^*(p,k)]    
    \sum_{\text{Pol.}}\sum_{r=1,2}\int \frac{d^4k_1}{(2\pi)^4} \frac{d^4k_2}{(2\pi)^4}   
      \notag  \\    & \times (2\pi)^4\delta^{(4)}(p'-k_1-k_2)  \nonumber \\
    &\times\left[\gamma^\mu u^{(r)}(k_2)\varepsilon_\mu(k_1)\varepsilon_\nu^{*}(k_1)\bar{u}^{(r)}(k_2)
    \gamma^\nu\right]\nonumber \\
&\times(2\pi)^2\delta(k_1^2)\delta  ^{(\text{phys)}}  (\Lambda_r^2(k_2^0))\theta(k_1^0)\theta(k_2^0)[J(p,k)]\,.
\end{align}
Finally, we recognize the amplitude for the cut diagram as
\begin{align}
i\mathcal{M}^{(2)}&=\left[\varepsilon_\nu^*(k_1)\bar{u}^r(k_2)(-ie\gamma^\nu)\right]\left( \frac{iF(p')}
{\Lambda_+^2(p')\Lambda_-^2(p')}\right)\nonumber \\
&\times \left[(-ie\gamma^\mu)u^s(p)\varepsilon_\mu(k)\right] \,.
\end{align}
We have the optical theorem satisfied 
for the one-loop Compton scattering diagram
\begin{align}
 2\text{Im}(\mathcal{M}_F ^{(2)})&=\sum_{\text{Pol.}}\sum_{r=1,2}\int\frac{d^4k_1}
 {(2\pi)^4}\frac{d^4k_2}{(2\pi)^4}(2\pi)^4    \\   &\times \delta^{(4)}(p'-k_1-k_2)\vert\mathcal{M}^{(2)}  \vert^2\nonumber \\
 &\times(2\pi)^2\delta(k_1^2)\delta  ^{(\text{phys)}} (\Lambda_r^2(k_2^0))\theta(k_1^0)\theta(k_2^0) \,. \notag
\end{align}
\section{Conclusions}\label{sec:V}
We have studied the unitarity of the $S$-matrix in a Lorentz-violating theory of modified 
QED with higher-order operators. As well known, higher-order operators in the Lagrangian density 
can lead to a potential loss of unitarity, especially in loop diagrams. 
The reason for this is that loop diagrams involve off-shell virtual particles, and so high energy modes 
associated with ghost states 
may, in principle, be propagated
through the cuts in the perturbative unitarity equation. This high contrasts with the situation 
where conservation of momentum selects only particles to be propagated in tree-level diagrams.

In our particular model, the selection of the preferred background in the pure timelike direction leads to
higher time derivatives and implies a negative metric sector. We have seen that 
the effective approach, together with the Lee-Wick prescription, can provide a unitarity theory.
 The Lee-Wick prescription is implemented by considering
only stable particles to have positive metric. 
A highlight of this work has been to provide a decoupled  unitarity equation restricted to the positive metric sector.
We have proved that 
the theory is unitary since no ghost modes are propagated through the cuts in the unitarity equation. 
A generalization of Cutkosky rule, at the one-loop order considered, is possible by introducing physical Dirac deltas defined to select only positive-metric solutions.
This extension can be useful for analyzing unitarity in higher-order diagrams.
\medskip
\acknowledgments
JLS thanks the Spanish Ministry of Universities and the European Union  
Next Generation EU/PRTR for the funds through the Maria 
Zambrano grant to attract international talent 2021 program. CMR 
has been funded by Fondecyt Regular No. 1191553, Chile. CR acknowledges 
support by ANID fellowship No. 21211384 from 
the Government of Chile and Universidad de Concepci\'on.

\end{document}